\documentclass[journal,twocolumn]{IEEEtran}

\usepackage[cmex10]{amsmath}
\usepackage[margin=0.9in]{geometry}

\usepackage{cite}

\usepackage{latexsym}
\usepackage{color}
\usepackage{graphics}
\usepackage{amssymb}
\usepackage{amsthm}
\usepackage{subcaption}

\usepackage{cite}
\usepackage{array}
\usepackage{graphicx}

\usepackage{url}

\newcommand{\argmax}[1]{{\underset{{#1}}{\mathrm{arg\,max}}}}

\newcommand{\vect}[1]{\mathbf{#1}}

\def\tr{\mathrm{tr}}

\def\Htran{\mbox{\tiny $\mathrm{H}$}}
\def\Ttran{\mbox{\tiny $\mathrm{T}$}}

\theoremstyle{definition}
\newtheorem{remark}{Remark}

\newtheorem{theorem}{Theorem}
\newtheorem{corollary}{Corollary}
\newtheorem{lemma}{Lemma}
\newtheorem{definition}{Definition}

\begin{document}

\title{A Random Access Protocol for Pilot Allocation in Crowded Massive MIMO Systems}

\IEEEoverridecommandlockouts

\author{Emil Bj{\"o}rnson,~\IEEEmembership{Member,~IEEE,}
        Elisabeth de Carvalho,~\IEEEmembership{Senior~Member,~IEEE,} \\
        Jesper H. S\o{}rensen,~\IEEEmembership{Member,~IEEE,}
        Erik G. Larsson,~\IEEEmembership{Fellow,~IEEE,}
        Petar Popovski,~\IEEEmembership{Fellow,~IEEE}%
        \thanks{\copyright \, 2017 IEEE. Personal use of this material is permitted. Permission from IEEE must be obtained for all other uses, in any current or future media, including reprinting/republishing this material for advertising or promotional purposes, creating new collective works, for resale or redistribution to servers or lists, or reuse of any copyrighted component of this work in other works.}%
\thanks{E.~Bj\"ornson and E.~G.~Larsson are with the Department of Electrical Engineering (ISY), Link\"{o}ping University, Link\"{o}ping, Sweden (email: emil.bjornson@liu.se, erik.g.larsson@liu.se).}%
\thanks{E.~de Carvalho, J.~H.~S\o{}rensen, and P.~Popovski are with the Department of Electronic Systems, Aalborg University, Aalborg, Denmark (email: edc@es.aau.dk, petarp@es.aau.dk).}%
\thanks{This work was presented in part at the IEEE International Conference on Communications (ICC), Kuala Lumpur, Malaysia, May 2016.}%
\thanks{This work was performed partly in the framework of the Danish Council for Independent Research (DFF133500273), the Horizon 2020 project FANTASTIC-5G (ICT-671660), the EU FP7 project MAMMOET (ICT-619086), ELLIIT, and CENIIT. The authors would like to acknowledge the contributions of the colleagues in FANTASTIC-5G and MAMMOET.}%
\thanks{Digital Object Identifier 10.1109/TWC.2017.2660489}%
}

\maketitle

\begin{abstract}
The Massive MIMO (multiple-input multiple-output) technology has great potential to manage the rapid growth of wireless data traffic. Massive MIMO achieves tremendous spectral efficiency by spatial multiplexing of many tens of user equipments (UEs). These gains are only achieved in practice if many more UEs can connect efficiently to the network than today. As the number of UEs increases, while each UE intermittently accesses the network, the random access functionality becomes essential to share the limited number of pilots among the UEs. In this paper, we revisit the random access problem in the Massive MIMO context and develop a reengineered protocol, termed \emph{strongest-user collision resolution} (SUCRe). An accessing UE asks for a dedicated pilot by sending an uncoordinated random access pilot, with a risk that other UEs send the same pilot.
The favorable propagation of Massive MIMO channels is utilized to enable distributed collision detection at each UE, thereby determining the strength of the contenders' signals and deciding to repeat the pilot if the UE judges that its signal at the receiver is the strongest. The SUCRe protocol resolves the vast majority of all pilot collisions in crowded urban scenarios and continues to admit UEs efficiently in overloaded networks.
\end{abstract}


\section{Introduction}
\label{sec:intro}

The number of wirelessly connected devices and their respective data traffic are growing rapidly as we transition into the fully networked society \cite{EricssonMobility}. The growth is currently driven by video streaming, social networking, and new use cases, such as machine-to-machine communication and internet of things. In dense urban scenarios, the METIS project predicts a future with up to 200.000 devices per km$^2$ and an associated data volume of 500 Gbyte/month/device \cite{METIS_D11_short}. These are massive numbers that call for radical changes in the network infrastructure to avoid congestion and guarantee service quality and availability.  A fair amount of the traffic is expected to be offloaded to WiFi and small-cell-technology operating at mm-wave frequencies, but macro cellular networks operating at frequencies of one or a few GHz  will remain to define the coverage, mobility support, and guaranteed service quality. Hence, future cellular networks need to handle urban deployment with \emph{massive} numbers of connected UEs that request \emph{massive} data volumes \cite{Boccardi2014a}. 

Since the cellular frequency bands are scarce, orders-of-magnitude capacity improvements are only possible by radically increasing the spectral efficiency (SE) [bit/s/Hz]. The Massive MIMO network topology, proposed in the seminal paper\cite{Marzetta2010a}, can theoretically deliver such extraordinary improvements \cite{Bjornson2016a}. The gains are achieved by spatial multiplexing, where base stations (BSs) with hundreds of antennas are utilized to serve tens of UEs per cell, at the same time-frequency resource. 
Massive MIMO is primarily a technology for time-division duplexing (TDD) \cite{Bjornson2016b}, where scalable channel estimation protocols are achieved by only requiring uplink pilots and utilizing channel reciprocity to obtain downlink channel estimates \cite{Vieira2014a}.
The communication-theoretic performance and asymptotic limits of Massive MIMO have been analyzed extensively in recent years; see for instance \cite{Marzetta2010a,Jose2011b,Huh2012a,Hoydis2013a,Ngo2013a,Yin2013a,Adhikary2013a,Bjornson2014a,Lu2014a,Bjornson2016a,Bjornson2016b}.
These works have focused on the performance achieved by \emph{active} UEs, mainly in homogeneously loaded cells, while the network access functionality has received little attention \cite{Bjornson2015d}. It is important to note that the building block of the traffic is data packets, which are generated in an intermittent and unpredictable fashion in all non-streaming applications. For example, web browsing and social network applications are characterized by bursty on-off traffic where intensive download activities are interwoven with long periods of silence. Many UEs thus switch between being active and inactive at a frequent and irregular basis.
Since Massive MIMO will be deployed to handle, say, $10\times$ higher UE densities than in current systems, the number of UEs that switch between active and inactive mode in a given time interval will be $10\times$ higher as well. Scalable and efficient access protocols are thus mandatory. 

\subsection{Random Access Functionality in LTE}

\begin{figure}[t!]
\begin{center}
\includegraphics[width=.9\columnwidth]{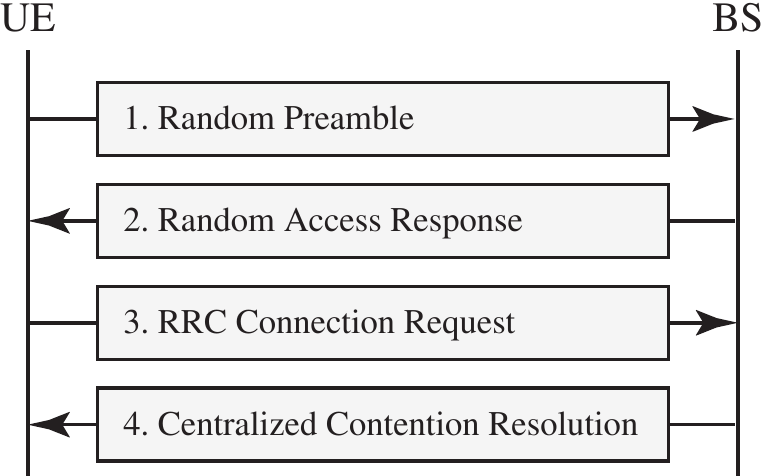}
\end{center} \vskip-3mm
\caption{The PRACH protocol of the LTE system.} \label{figure:PRACH}  \vskip-3mm
\end{figure}

Before we propose a new highly scalable access protocol, we describe the conventional protocol used on the Physical Random Access Channel (PRACH) of Long-Term Evolution (LTE). It consists of four steps \cite{Hasan2013a}, as illustrated in Fig.~\ref{figure:PRACH}. In Step~1, each accessing UE picks a preamble at random from a predefined set. The preamble is an entity that enables synchronization towards the BS. It does not carry specific reservation information or data and thus can be viewed as a pilot sequence. Since UEs that wish to access the network are not coordinated in picking the preambles, a collision occurs if two or more UEs select the same preamble simultaneously. A BS in LTE only detects if a specific preamble is active or not in Step~1 \cite{CodeExpanded}. In Step~2, the BS sends a \emph{random access response} corresponding to each activated preamble, to convey physical parameters (e.g., timing advance) and allocate a resource to the UE (or UEs) that activated the preamble.
In Step 3, each UE that has received a response to its preamble transmission sends a \emph{RRC (Radio Resource Control) Connection Request} in order to request resources for subsequent data transmission. If more than one UE activated the preamble, then all these UEs use the same resource to send their RRC connection request in Step 3 and this collision is detected by the BS. Step 4 is called \emph{contention resolution} and contains one or multiple steps to resolve the collision. This is a complicated procedure that might result in that all colliding UEs need to make a new access attempt after a random waiting time.

\subsection{Prior Work on Random Access in Massive MIMO}

Conventional cellular networks allocate dedicated resources to each active UE, thus the BS needs to convey the time-frequency positions of these resources. In contrast, Massive MIMO systems allocate all time-frequency resources to all active UEs, and separate them spatially based on their pilot sequences. The number of pilots is limited by the size of the channel coherence block. In the original Massive MIMO concept of\cite{Marzetta2010a} the UEs within a cell use mutually orthogonal pilots, while the necessary reuse of pilots across cells causes \emph{inter-cell pilot contamination} that leads to additional interference \cite{Jose2011b}.

In crowded urban scenarios, the total number of active and inactive UEs residing in a cell is also much larger than the number of available pilot sequences. Hence, a pilot cannot be permanently associated with a UE in the cell but need to be opportunistically allocated and deallocated to follow its unpredictable intermittent data traffic pattern.  Random access (RA) mechanisms can be used for such pilot allocation, but must be designed to resolve collisions in crowded cells without excessive access delays. The papers \cite{Sorensen2014a,deCarvalho2016a,Sanguinetti2016a} propose protocols where UEs  can transmit data whenever they like by selecting pilots at random from a common pool. This eliminates access delays, at the cost of pilot collisions that cause \emph{intra-cell pilot contamination}. The collisions are expressed as a graph code in \cite{Sorensen2014a} and belief propagation is used to mitigate pilot contamination, by utilizing the principles of coded random access \cite{Paolini2011a,PSLP2014}. Several SE expressions are derived in \cite{deCarvalho2016a} and utilized to optimize the UE activation probability and pilot length. Asynchronous UE timings are utilized in \cite{Sanguinetti2016a} to detect and resolve pilot collisions in RA.

In this paper, we explore the alternative solution that each UE needs to be assigned a pilot sequence before transmitting payload data, to avoid intra-cell pilot collisions and actualize the payload transmission situation considered in the main body of Massive MIMO research \cite{Marzetta2010a,Jose2011b,Huh2012a,Hoydis2013a,Ngo2013a,Bjornson2016a,Bjornson2016b}. We focus on urban deployments with small initial timing variations and propose a new RA protocol for UEs that wish to access the network. The protocol can resolve RA collisions in a distributed and scalable way, by exploiting special properties of Massive MIMO channels. A preliminary version of the protocol was described in \cite{Bjornson2016e}.

\subsection{Outline and Notation}

The remainder of this paper is organized as follows. Section~\ref{sec:proposed-protocol} describes the proposed RA protocol assuming arbitrary Massive MIMO channels, while Section~\ref{sec:Rayleigh-fading} analyzes the performance with uncorrelated Rayleigh fading channels.
Section~\ref{section:numerical-results} provides numerical results that show how the proposed protocol can resolve severe RA collisions and operate under very high load. The main conclusions are summarized in Section~\ref{sec:conclusion}.

The following notation is used throughout the paper. The transpose, conjugate-transpose, and conjugate of a matrix $\vect{X}$ are denoted by $\vect{X}^{\Ttran}$, $\vect{X}^{\Htran}$, and $\vect{X}^{*}$, respectively.
We let $\vect{I}_{M}$ denote the $M \times M$ identity matrix, whereas $\| \cdot \|$ 
and $| \cdot |$ stand for the Euclidean norm of a vector and cardinality of a set, respectively. The notations $\mathbb{E}\{ \cdot \}$ and $\mathbb{V}\{ \cdot \}$  indicate the expectation and variance with respect to a random variable. We use $\mathcal{N}(\cdot,\cdot)$ to denote a Gaussian distribution, $\mathcal{CN}(\cdot,\cdot)$ to denote a  circularly-symmetric complex Gaussian distribution, $B(\cdot,\cdot)$ to denote a binomial distribution, and $\chi_n$ to denote a chi-distribution with $n$ degrees of freedom. We use $ \mathbb{C}$  and $\mathbb{R}$ to denote spaces of complex-valued and real-valued numbers, respectively. The Gamma function is denoted by $\Gamma(\cdot)$.

\begin{figure*}[t!]
\begin{center}
\includegraphics[width=2\columnwidth]{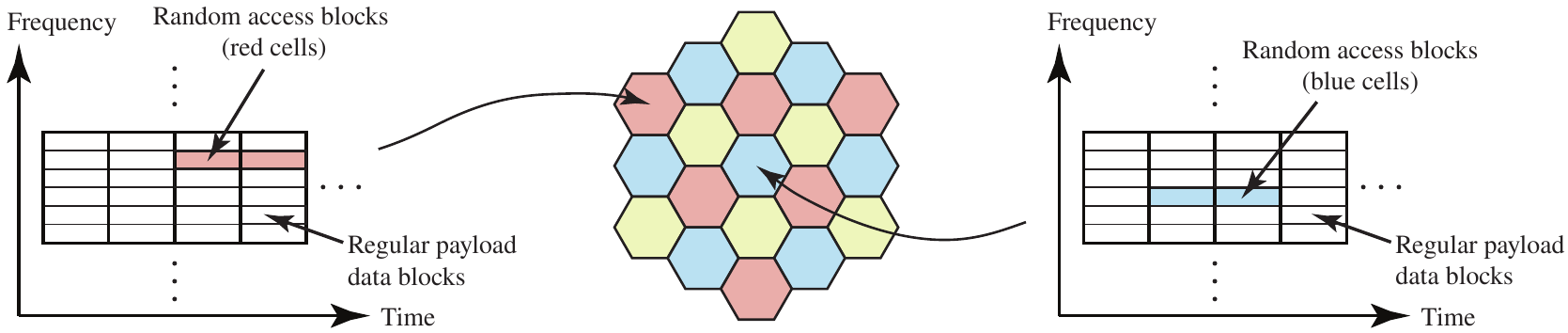}
\end{center} \vskip-3mm
\caption{Illustration of the proposed transmission protocol, where the time-frequency domain is divided into coherence blocks. The majority of the  blocks are used for payload data transmission for active UEs (which have been allocated dedicated pilots). A few of the blocks are used for random access where inactive UEs can ask to be admitted to the payload data blocks. The location of the random access blocks in the time-frequency grid is different between adjacent cells (here illustrated using a reuse factor of three), and the neighbors are either quiet or send payload during the RA.} \label{figure:block-structure-multi-cell} \vskip-3mm
\end{figure*}

\section{Proposed Random Access Protocol}
\label{sec:proposed-protocol}

We consider cellular networks where each BS is equipped with $M$ antennas. The system operates in TDD mode and the time-frequency resources are divided into coherence blocks of $T$ channel uses, dimensioned such that the channel responses between each BS and its UEs are constant and frequency flat within a block, while they vary between blocks. This can be implemented using orthogonal frequency-division multiplexing (OFDM). We let $\mathcal{U}_i$ denote the set of UEs that reside in cell~$i$. At any given time, only a subset $\mathcal{A}_i \subset \mathcal{U}_i$ of the UEs are active in the sense that they are transmitting and/or receiving data.
Note that in the scenarios relevant for Massive MIMO deployment, we typically have a very large UE set: $|\mathcal{U}_i| \gg T$. However, the active UEs satisfy $| \mathcal{A}_i| < T$, thus the BS can temporarily assign orthogonal pilot sequences to these UEs and reclaim them when their respective transmissions are finished.

The coherence blocks are divided into two categories: \emph{payload data blocks} and \emph{random access blocks}. For each cell~$i$, 
the first category is used for uplink (UL) and downlink (DL) data transmission to the UEs in the set $\mathcal{A}_i$. These UEs have temporarily been allocated $|\mathcal{A}_i |$ mutually orthogonal pilot sequences, which however are reused in other cells (using some reuse factor that guarantees low pilot contamination \cite{Bjornson2016a}). The payload data blocks can be operated as in the classical Massive MIMO works \cite{Marzetta2010a,Jose2011b,Huh2012a,Hoydis2013a,Ngo2013a,Bjornson2016a,Bjornson2016b}, which provide methods to achieve high data rates.

The second category is dedicated for RA from inactive UEs (i.e., some of those in $\mathcal{U}_i \! \setminus \!  \mathcal{A}_i$) that wish to be admitted to the payload data blocks; that is, to be allocated a temporary dedicated pilot.
This category has not been studied in the Massive MIMO context and is the main focus of this paper. The time-frequency location of the RA blocks is different between adjacent cells, as illustrated in Fig.~\ref{figure:block-structure-multi-cell}. This design choice protects the RA procedure from the strongest forms of inter-cell interference, as will be shown later.

\vspace{-4mm}

\subsection{Strongest-user collision resolution (SUCRe)---Overview}

\begin{figure}[t!]
\begin{center}
\includegraphics[width=.9\columnwidth]{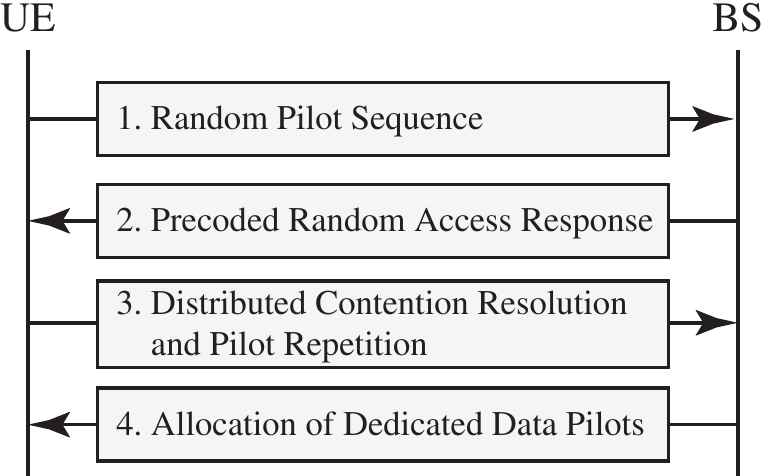}
\vspace{-0.3cm}
\end{center}
\caption{The proposed SUCRe random access protocol for Massive MIMO.} \label{figure:proposed-protocol}
\end{figure}

The key contribution of this paper is the strongest-user collision resolution (SUCRe) protocol, which is an efficient way to operate the RA blocks in beyond-LTE Massive MIMO systems. We first give a brief overview of the protocol and then provide the exact analytical details. The four main steps of the SUCRe protocol are illustrated in Fig.~\ref{figure:proposed-protocol}. There is also a preliminary Step~0 in which the BS broadcasts a control signal. Each UE uses this signal to estimate its average channel gain and to synchronize itself towards the BS. In OFDM, the UE and the BS need to be synchronized in frequency and the timing delay can be neglected if it is shorter than the cyclic prefix. The round-trip time determines the maximum timing delay, thus the normal CP in LTE allows for 750 m cell radius and the extended CP allows for 2.5 km---these are substantially larger than the 250--500 m cell radius typical in urban deployments. This paper focuses on such urban scenarios and we stress that the spatial multiplexing in Massive MIMO is ideal for crowded urban settings.
In Step 1, a subset of the inactive UEs in cell~$i$ wants to become active. Each such UE selects a pilot sequence at random from a predefined pool of RA pilots. BS~$i$ estimates the channel that each pilot has propagated over.  If multiple UEs selected the same RA pilot, a collision has occurred and the BS obtains an estimate of the superposition of the UE channels.
The BS cannot detect if collisions occurred at this point, which resembles the situation in LTE.

\begin{figure}[t!]
\begin{center}
\includegraphics[width=8cm]{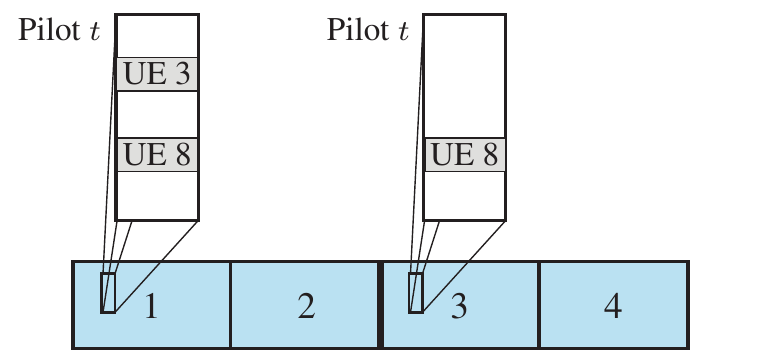}
\vspace{-0.3cm}
\end{center}
\caption{Illustration of a UE collision at RA pilot $t$ resolved by the SUCRe protocol, where only the UE with strongest signal gain repeats the pilot in Step 3.} \label{figure:proposed-protocol2}
\end{figure}

In Step 2, the BS responds by sending DL pilots that are precoded using the channel estimates, which results in spatially directed signals toward the UEs that sent the particular RA pilot. The DL signal features an array gain of $M$ that is divided between the UEs that sent the RA pilot. Due to channel reciprocity, the share of the array gain is proportional to their respective UL signal gains, particularly when $M$ is large, which enables each UE to estimate the sum of the signal gains and compare it with its own signal gain (using information obtained in Step 0). Each UE can thereby detect RA collisions in a \emph{distributed way}. This departs from the conventional approach in which collisions are detected in a \emph{centralized way} at the BS and broadcasted to the UEs.

Based on the detection in Step~2 and the favorable propagation of Massive MIMO channels, the UEs can resolve RA contentions in Step~3 by applying a local decision rule: only the UE with the strongest UL signal gain should repeat the RA pilot. This is a key advantage over LTE, where all contending UEs repeat the preamble in Step~3. This mechanism is illustrated in Fig.~\ref{figure:proposed-protocol2}. The probability of non-colliding pilot transmission in Step~3 is vastly increased in the SUCRe protocol, which enables the network to admit UEs also in crowded scenarios. The transmission in Step~3 also contains the identity of the UE and a request for payload transmission, resembling the RRC Connection Request in LTE. Step 4 grants these resources by assigning a pilot sequence that can be used in the payload blocks or starts further contention resolution (e.g., in an LTE fashion or by repeating the SUCRe protocol) in the few cases when RA collisions remain. Hence, the SUCRe protocol both stands on its own and can complement conventional contention resolution methods.

\subsection{Detailed Description of the SUCRe Protocol}

Next, we describe and analyze the proposed RA protocol in detail. Without loss of generality, we focus on an arbitrary cell, say cell~$0$, and consider how interference from other cells impacts the operation. Let $\mathcal{K}_0 = \mathcal{U}_0 \! \setminus \!  \mathcal{A}_0$ denote the set of inactive UEs in cell~$0$ in a given RA block. Hence, there are $K_0 = |\mathcal{K}_0|$ inactive UEs in cell~$0$. These are assumed to share $\tau_p$ mutually orthogonal RA pilot sequences $\boldsymbol{\psi}_1,\ldots,\boldsymbol{\psi}_{\tau_p} \in \mathbb{C}^{\tau_p}$ that span $\tau_p$ UL channel uses and  satisfy $\| \boldsymbol{\psi}_{t} \|^2=\tau_p$. The inactive UEs are not fully time-synchronized, but the pilot orthogonality is  maintained at the receiver since we consider urban scenarios where the roundtrip delays are smaller than the cyclic prefix. We typically have $\tau_p \ll K_0$, but there is no formal constraint. 

Each of the $K_0$ UEs picks one of the $\tau_p$ pilots uniformly at random in each RA block: UE~$k$ selects pilot $c(k) \in \{ 1, 2, \ldots , \tau_p\}$.
Furthermore, each UE would like to become active in the current block with probability $P_a \leq 1$, which is a fixed scenario-dependent parameter that describes how often a UE has data packets to transmit or receive. An access attempt by UE~$k$ consists of transmitting the pilot $\boldsymbol{\psi}_{c(k)}$ with a non-zero power $\rho_k>0$, otherwise it stays silent by setting $\rho_k=0$.
Hence, each inactive UE will transmit a particular pilot sequence $\boldsymbol{\psi}_t$ (using non-zero power) with probability $P_a / \tau_p$. The set $\mathcal{S}_t = \{k: c(k) = t, \rho_k > 0 \}$ contains the indices of the UEs that transmit pilot $t$. Based on this model, the number of UEs, $| \mathcal{S}_t |$, that transmits  $\boldsymbol{\psi}_t$ has a binomial distribution:\footnote{This is a simplified model where each RA block is treated independently. However, a UE that fails to access the network in one block will soon try again, which creates a correlation between RA blocks. This practical scenario is studied numerically in Section~\ref{section:numerical-results} and we stress that the proposed protocol can be applied under any UE distribution.}
\begin{equation} \label{eq:binomial-dist-N}
| \mathcal{S}_t | \sim B \left( K_0 , \frac{P_a}{\tau_p} \right). 
\end{equation}
We notice that pilot $t$ is unused ($| \mathcal{S}_t |=0$)  with probability $(1-\frac{P_a}{\tau_p})^{K_0}$ and selected by only one  UE ($| \mathcal{S}_t |=1$) with probability $K_0 \frac{P_a}{\tau_p} (1-\frac{P_a}{\tau_p})^{K_0-1}$. Consequently, an RA collision ($| \mathcal{S}_t | \geq 2$) occurs at this arbitrary pilot with probability
\begin{equation}
1 - \left( 1-\frac{P_a}{\tau_p} \right)^{K_0} - K_0 \frac{P_a}{\tau_p} \left(1-\frac{P_a}{\tau_p} \right)^{K_0-1}.
\end{equation} 
These collisions need to be detected and resolved before any UE can be admitted into the payload blocks. The SUCRe protocol is a distributed method to resolve pilot collisions at the UE side by utilizing properties of Massive MIMO channels. 

The channel vector between UE~$k \in \mathcal{K}_0$ and its BS is denoted by $\vect{h}_k \in \mathbb{C}^M$. We adopt a very general propagation model where the channels are assumed to satisfy the following two conditions (almost surely):
\begin{align} \label{eq:favorable-1}
\frac{\| \vect{h}_k \|^2}{M} &\xrightarrow{M \rightarrow \infty} \beta_k, \quad \forall k, \\
\frac{\vect{h}_k^{\Htran} \vect{h}_i}{M} &\xrightarrow{M \rightarrow \infty} 0, \quad \forall k,i, \,\, k \neq i, \label{eq:favorable-2}
\end{align}
for some strictly positive value of $\beta_k$ that is known to UE~$k$ (it was estimated in Step 0). Such channels are said to offer \emph{channel hardening} and \emph{asymptotic favorable propagation}~\cite{Ngo2014a}.

The properties \eqref{eq:favorable-1} and \eqref{eq:favorable-2} are satisfied (almost surely) by a variety of stochastic channel models; for example, when $\vect{h}_k = \vect{R}_k^{1/2} \vect{x}_k$ where $\vect{R}_k \in \mathbb{C}^{M \times M}$ is a positive semi-definite matrix with bounded spectral norm and $\vect{x}_k \in \mathbb{C}^M$ has i.i.d. entries with zero mean and bounded eighth-order moment \cite[Theorem 3.4, Theorem 3.7]{Couillet2011a}. In this case we have $\tr(\vect{R}_k)/M \rightarrow \beta_k$. Asymptotic favorable propagation can also be obtained for deterministic line-of-sight channels; for example, for uniform linear arrays (ULAs) where the UEs have distinct angles with respect to the BS array \cite{Ngo2014a}.

We will describe the four steps of the SUCRe protocol under the assumption that the channels satisfy \eqref{eq:favorable-1} and \eqref{eq:favorable-2}. In Section~\ref{sec:Rayleigh-fading}, we particularize the protocol for uncorrelated Rayleigh fading channels, which enables us to quantify the performance in further detail.

\subsubsection{Random Pilot Sequence}

In Step 1 of the proposed RA protocol, the BS receives the signal $\vect{Y} \in \mathbb{C}^{M \times \tau_p}$ from the pilot transmission:
\begin{equation}
\vect{Y} = \sum_{k \in \mathcal{K}_0} \sqrt{\rho_k} \vect{h}_k \boldsymbol{\psi}_{c(k)}^{\Ttran} + \vect{W} + \vect{N},
\end{equation}
where $\vect{N} \in \mathbb{C}^{M \times \tau_p}$ is independent receiver noise with each element distributed as $\mathcal{CN}(0, \sigma^2  )$.
The matrix $\vect{W} \in \mathbb{C}^{M \times \tau_p}$ represents interference from other cells.

By correlating $\vect{Y}$ with an arbitrary (normalized) pilot sequence $\boldsymbol{\psi}_t$, the BS obtains
\begin{align} \notag
\vect{y}_t = \vect{Y} \frac{\boldsymbol{\psi}_t^*}{\| \boldsymbol{\psi}_t \|} &=  \sum_{i \in \mathcal{S}_t} \sqrt{\rho_i} \| \boldsymbol{\psi}_t \| \vect{h}_i +  \vect{W} \frac{\boldsymbol{\psi}_t^*}{\| \boldsymbol{\psi}_t \|}+ \vect{n}_t \\ &
=  \sum_{i \in \mathcal{S}_t} \sqrt{\rho_i \tau_p}  \vect{h}_i +  \vect{W} \frac{\boldsymbol{\psi}_t^*}{\| \boldsymbol{\psi}_t \|}+ \vect{n}_t, \label{eq:yt}
\end{align}
where $\vect{n}_t = \vect{N} \frac{\boldsymbol{\psi}_t^*}{\| \boldsymbol{\psi}_t \|} \sim \mathcal{CN}(\vect{0}, \sigma^2 \vect{I}_M)$ is the effective receiver noise and we notice that $\| \boldsymbol{\psi}_t \| = \sqrt{\tau_p}$.
Recall that $\mathcal{S}_t$ is the set of UEs that transmitted pilot $\boldsymbol{\psi}_t$.

The inter-cell interference $\vect{W}$ can be modeled as
\begin{equation} \label{eq:intercell-interference-term}
\vect{W} = \sum_l \vect{w}_l \vect{d}_l^{\Ttran} + \sum_{t=1}^{\tau_p} \sum_{k \in \mathcal{S}_t^{\textrm{interf}}} \sqrt{\rho_{t,k}} \vect{g}_{t,k} \boldsymbol{\psi}_{t}^{\Ttran}.
\end{equation}
The first summation in \eqref{eq:intercell-interference-term} is over the interfering data transmissions carried out in neighboring cells with another color than cell~$0$ in Fig.~\ref{figure:block-structure-multi-cell}. The $l$th interferer has the channel $\vect{w}_l \in \mathbb{C}^M$ to the BS in cell~$0$ and transmits some random data sequence $\vect{d}_l \in \mathbb{C}^{\tau_p}$.
The second summation is over the interferers in cells with the same color as cell~$0$ in Fig.~\ref{figure:block-structure-multi-cell}, which also perform RA. The interferers that use pilot $\boldsymbol{\psi}_{t}$ are gathered in the set $\mathcal{S}_t^{\textrm{interf}}$, and member $k \in \mathcal{S}_t^{\textrm{interf}}$ has the channel  $\vect{g}_{t,k}$ to the BS in cell~$0$ and uses the transmit power $\rho_{t,k}$.
 It follows that 
\begin{equation} 
\vect{W} \frac{\boldsymbol{\psi}_t^*}{\| \boldsymbol{\psi}_t \|} = 
\sum_l \vect{w}_l  \frac{\vect{d}_l^{\Ttran}\boldsymbol{\psi}_t^*}{\| \boldsymbol{\psi}_t \|} +  \sum_{k \in \mathcal{S}_t^{\textrm{interf}}} \sqrt{\rho_{t,k} \tau_p} \vect{g}_{t,k}.
\end{equation}
Assuming that all the interfering channels also satisfy the conditions in \eqref{eq:favorable-1} and \eqref{eq:favorable-2}, denoted as $\| \vect{w}_l \|^2 /M \rightarrow \beta_{w,l}$ and $\| \vect{g}_{t,k} \|^2/M \rightarrow \beta_{t,k}$, we obtain
\begin{equation}
\frac{ \left| \vect{W} \frac{\boldsymbol{\psi}_t^*}{\| \boldsymbol{\psi}_t \|} \right|^2}{M}  \rightarrow  
\underbrace{\sum_l \beta_{w,l}  \frac{|\vect{d}_l^{\Ttran}\boldsymbol{\psi}_t^*|^2}{\| \boldsymbol{\psi}_t \|^2}  + \! \sum_{k \in \mathcal{S}_t^{\textrm{interf}}} \rho_{t,k} \tau_p \beta_{t,k}}_{\omega_t}
\end{equation}
as $M \rightarrow \infty$. Note that there is interference in $\omega_t$ from both data transmission and RA pilots in other cells, but the former typically dominates since the closest neighboring cells in Fig.~\ref{figure:block-structure-multi-cell} transmits data. It is only the value of $\omega_t$ that is important in the remainder of this section, and not how it is computed.

\begin{remark}[Detecting active pilots]
The BS can utilize $\vect{y}_t \in \mathbb{C}^{M}$ to determine if $| \mathcal{S}_t | \geq 1$ or $| \mathcal{S}_t | = 0$ for the considered pilot sequence; that is, whether or not there is at least one active UE. 
This is particularly easy in Massive MIMO systems that operate under channel hardening and asymptotic favorable propagation since
\begin{equation} \label{eq:detect-usage}
\frac{\| \vect{y}_t  \|^2}{M} \xrightarrow{M \rightarrow \infty}  \sum_{i \in \mathcal{S}_t} \rho_i  \beta_i \tau_p +  \omega_t + \sigma^2.
\end{equation}
This is an additional feature since the SUCRe protocol does not require the BS to know which pilots were used in Step 1.
\end{remark}

\subsubsection{Precoded Random Access Response}

In Step 2, the BS responds to the RA pilots by sending orthogonal precoded DL pilot signals that correspond to each of the RA pilots that were used in the UL. 
The response to $\boldsymbol{\psi}_t$ is a pilot sequence $\boldsymbol{\phi}_t \in \mathbb{C}^{\tau_p}$, and the DL pilot sequences $\boldsymbol{\phi}_1,\ldots,\boldsymbol{\phi}_{\tau_p} \in \mathbb{C}^{\tau_p}$ are mutually orthogonal and satisfy $\| \boldsymbol{\phi}_{t} \|^2=\tau_p$.
Note that  $\vect{y}_t$ in \eqref{eq:yt} is a weighted superposition of the channels of UEs that used pilot $t$ (it is also impaired by interference and noise). If the BS uses the normalized conjugate $\vect{y}_t^* / \| \vect{y}_t \|$ as precoding vector when sending the pilot sequence $\boldsymbol{\phi}_t$ in the DL, the signal will be directed in a multi-cast maximum ratio transmission fashion towards the UEs in $\mathcal{S}_t$.
The complete precoded DL pilot signal $\vect{V} \in \mathbb{C}^{M \times \tau_p}$ is 
\begin{equation} \label{eq:DL-precoded-pilot}
\vect{V} =  \sqrt{q} \sum_{t=1}^{\tau_p}  \frac{\vect{y}_t^*}{\| \vect{y}_t \|} \boldsymbol{\phi}_t^{\Ttran},
\end{equation}
where the DL transmit power $q$ has a predefined value. Note that all pilot sequences are also sent in the DL and the pilot length $\tau_p$ is independent of the number of antennas.

The received signal $\vect{z}_k \in \mathbb{C}^{\tau_p}$ at UE~$k \in \mathcal{S}_t$ is
\begin{equation} \label{eq:received-user-k}
\vect{z}_k^{\Ttran} = \vect{h}_k^{\Ttran} \vect{V} + \boldsymbol{\upsilon}_k^{\Ttran} + \boldsymbol{\eta}_k^{\Ttran},
\end{equation}
where $\vect{h}_k^{\Ttran}$ is the reciprocal DL channel, $\boldsymbol{\upsilon}_k \in \mathbb{C}^{\tau_p}$ is inter-cell interference, and $\boldsymbol{\eta}_k \sim \mathcal{CN}(\vect{0},\sigma^2 \vect{I}_{\tau_p})$ is receiver noise.
By correlating the received signal $\vect{z}_k$ with the (normalized) DL pilot sequence $\boldsymbol{\phi}_t$, the UE obtains
\begin{equation} \label{eq:received-user-k-scalar}
z_k = \vect{z}_k^{\Ttran} \frac{\boldsymbol{\phi}_t^*}{\| \boldsymbol{\phi}_t \|}= \sqrt{q \tau_p}  \vect{h}_k^{\Ttran} \frac{\vect{y}_t^*}{\| \vect{y}_t \|} + \boldsymbol{\upsilon}_k^{\Ttran} \frac{\boldsymbol{\phi}_t^*}{\| \boldsymbol{\phi}_t \|} + \eta_k
\end{equation}
where $\eta_k = \boldsymbol{\eta}_k^{\Ttran} \frac{\boldsymbol{\phi}_t^*}{\| \boldsymbol{\phi}_t \|} \sim \mathcal{CN}(0,\sigma^2)$ is the effective receiver noise.
We notice that
\begin{align} \notag 
\frac{z_k}{\sqrt{M}} &= 
\sqrt{q \tau_p}   \frac{(\vect{h}_k^{\Htran} \vect{y}_t)^*}{M} \frac{1}{\sqrt{\frac{1}{M} \| \vect{y}_t \|^2 }} +  \frac{\boldsymbol{\upsilon}_k^{\Ttran} \boldsymbol{\phi}_t^*}{\sqrt{M} \| \boldsymbol{\phi}_t \|} + \frac{\eta_k}{\sqrt{M}} \\
& \xrightarrow{M \rightarrow \infty}  \frac{\sqrt{\rho_k q }  \beta_k \tau_p}{\sqrt{\sum_{i \in \mathcal{S}_t} \rho_i \beta_i \tau_p +  \omega_t + \sigma^2}} \label{eq:asymptotics-received-DL-channel}
\end{align}
by utilizing the asymptotic favorable propagation, the convergence in \eqref{eq:detect-usage}, exploiting the fact that the noise does not increase with $M$, and assuming that the inter-cell interference $\boldsymbol{\upsilon}_k$ is unaffected by $M$. The latter assumption is well motivated if the closest interfering cells that send RA pilots assign the DL pilots in different ways, to avoid causing coherent pilot contaminated interference.\footnote{For example, one cell might assign $\boldsymbol{\phi}_t=\boldsymbol{\psi}_t$, the next one assigns $\boldsymbol{\phi}_t=\boldsymbol{\psi}_{t+1}$, and another one assigns $\boldsymbol{\phi}_t=\boldsymbol{\psi}_{t+2}$. With such pilot switching, the closest $\tau_p$ cells that perform RA will not have any UEs that collide in both the UL and DL, which alleviates the coherent pilot contaminated interference between these cells.}

Looking at \eqref{eq:asymptotics-received-DL-channel}, let us define
\begin{equation}
\alpha_t = \sum_{i \in \mathcal{S}_t} \rho_i  \beta_i \tau_p +  \omega_t
\end{equation}
as the sum of the signal and interference gains received at the BS during the UL transmission of pilot $\boldsymbol{\psi}_t$. The intra-cell signals are amplified by a factor $\tau_p$, as compared to the inter-cell interference, which is the processing gain from having a pilot sequence that spans $\tau_p$ symbols. Let the function $\Re(\cdot)$ give the real part of its input. Based on \eqref{eq:asymptotics-received-DL-channel}, we obtain the approximation 
\begin{equation} 
\begin{split} 
\frac{\Re( z_k )}{\sqrt{M}} \approx \frac{\sqrt{\rho_k q }  \beta_k \tau_p}{\sqrt{\alpha_t +  \sigma^2}},
 \end{split}
\end{equation}
where we discard the imaginary part of $z_k$ that only contains noise, interference and estimation errors. UE~$k$ can use this approximation to  estimate $\alpha_t$:
\begin{equation} \label{eq:alpha-approx-est}
\hat{\alpha}_{t,k}^{\text{approx1}} = \max \left(
\frac{M q \rho_k  \beta_k^2 \tau_p^2 }{ \left( \Re ( z_k ) \right)^2 } - \sigma^2 , \, \rho_k  \beta_k \tau_p  \right),
\end{equation}
where $\max(\cdot,\cdot)$ takes the maximum of the two values (since UE~$k$ knows that $\alpha_t \geq \rho_k  \beta_k \tau_p$).
This estimator is asymptotically error-free, as $M \rightarrow \infty$, due to \eqref{eq:asymptotics-received-DL-channel}.

\subsubsection{Contention Resolution \& Pilot Repetition}

The UL pilot transmission is repeated in Step 3. The main goal of the proposed protocol is to resolve pilot contentions in a distributed manner in Step 3, so that each pilot is only repeated by one UE. Each UE~$k \in \mathcal{S}_t$ knows its own average signal gain $\rho_k \beta_k \tau_p$ and has an estimate $\hat{\alpha}_{t,k}$ of the sum of the signal gains of the contending UEs (plus inter-cell interference), such that it can infer:
\begin{itemize}
\item If a pilot collision has occurred: $\hat{\alpha}_{t,k} > \rho_k \beta_k \tau_p$ (with a margin that accounts for inter-cell interference);
\item How strong its own signal is relative to the sum of all the contenders' signals: $\rho_k \beta_k \tau_p / \hat{\alpha}_{t,k}$.
\end{itemize}
Since the number of contenders, $| \mathcal{S}_t |$, is unknown, a UE can only compare its own signal gain with the summation of the gains of its contenders. To resolve the contention we make the following definition.

\begin{definition}
The contention winner is the UE~$k \in \mathcal{S}_t$ with the largest $\rho_k \beta_k \tau_p$, referred to as the \emph{strongest user}.
\end{definition}

In the asymptotic case when $\alpha_t$ is exactly known, UE~$k$ is sure to be the contention winner if $\rho_k \beta_k \tau_p > \alpha_t - \rho_k\beta_k \tau_p$, irrespective of how many contenders there are. This criterion can be written as $\rho_k \beta_k \tau_p > \alpha_t/2$ and interpreted as having a UE~$k$ with a signal gain that is greater than the sum of all the other UEs' signal gains.

\begin{definition}
We have \emph{resolved a collision} if and only if a single UE appoints itself the contention winner.
\end{definition}

An example where a two-UE collision is resolved is given in Fig.~\ref{figure:proposed-protocol2}. If there are  $|\mathcal{S}_t|=2$ UEs (or $|\mathcal{S}_t|=1$ for that matter) the criterion $\rho_k \beta_k \tau_p > \alpha_t/2$ can be used to resolve any such collision, in the special case when $\alpha_t$ is known and there is no inter-cell interference. In practice, only the estimate $\hat{\alpha}_{t,k}$ is known, it happens that $|\mathcal{S}_t| \geq 3$ and there will be inter-cell interference. There is then a risk that multiple UEs or no UE identify themselves as the contention winner.

\begin{definition}
A \emph{false negative} occurs when none of the colliding UEs identifies itself as the contention winner. A \emph{false positive} occurs when more than one colliding UE identify itself as the contention winner.
\end{definition}

We propose that each UE~$k \in \mathcal{S}_t$ applies the following distributed \emph{decision rule}:
\begin{align} \label{eq:retransmission-rule}
&\mathcal{R}_{k}:  \quad \rho_k \beta_k \tau_p > \hat{\alpha}_{t,k} /2 + \epsilon_k \quad \textrm{(repeat)}, \\
&\mathcal{I}_{k}: \quad \rho_k \beta_k \tau_p \leq \hat{\alpha}_{t,k} /2 + \epsilon_k \quad \textrm{(inactive)}.
\end{align}
UE~$k \in \mathcal{S}_t$ concludes that it has the strongest signal gain if $\mathcal{R}_{k}$ is true and \emph{repeats} the transmission of pilot $\boldsymbol{\psi}_t$ in Step 3. If it instead concludes that $\mathcal{I}_{k}$ is true, it decides to remain \emph{inactive} by pulling out from the RA attempt and try again later. The estimation errors and inter-cell interference can cause false positives or negatives. The bias parameter $\epsilon_k \in \mathbb{R}$ can be used to tune the system behavior to the final performance criterion; for example, to maximize the average number of resolved collisions or to minimize the risk of false positives (or negatives).

The probability of resolving a contention is determined by the decision rule. By numbering the active UEs in $\mathcal{S}_t$ from $1$ to $| \mathcal{S}_t|$, the probability of resolving a pilot collision with $|\mathcal{S}_t|$ contenders is
\begin{align} \notag
P_{|\mathcal{S}_t|,\textrm{resolved}} &= 
\mathrm{Pr} \{ \mathcal{R}_{1},\mathcal{I}_{2},\ldots,\mathcal{I}_{| \mathcal{S}_t|} \} \\ &+  \notag
\mathrm{Pr} \{ \mathcal{I}_{1},\mathcal{R}_{2},\mathcal{I}_{3},\ldots,\mathcal{I}_{| \mathcal{S}_t|} \} + \ldots \\ &+ 
\mathrm{Pr} \{ \mathcal{I}_{1},\ldots,\mathcal{I}_{| \mathcal{S}_t|-1},\mathcal{R}_{| \mathcal{S}_t|} \},
 \label{eq:prob-resolved}
\end{align}
where the randomness is due to channel realizations, inter-cell interference, and noise (and possibly also random UE locations).
In the special case $|\mathcal{S}_t|=2$, \eqref{eq:prob-resolved} reduces to
\begin{equation} \label{eq:prob-resolved-S2}
\begin{split}
P_{2,\textrm{resolved}} &= 
\mathrm{Pr} \{ \mathcal{R}_{1},\mathcal{I}_{2} \} +  
\mathrm{Pr} \{ \mathcal{I}_{1},\mathcal{R}_{2} \},
\end{split}
\end{equation}
while a false negative occurs if both the UEs pull out (with probability $\mathrm{Pr} \{ \mathcal{I}_{1},\mathcal{I}_{2} \}$) and a false positive occurs when both UEs repeat the pilot (with probability $\mathrm{Pr} \{ \mathcal{R}_{1},\mathcal{R}_{2} \}$).

\begin{remark}[Probabilistic bias terms] \label{remark:bias-terms}
The decision rule in \eqref{eq:retransmission-rule} appears to make a hard decision on whether or not UE~$k$ is the strongest user. This can, however, be softened  by using random bias terms. For example, UE~$k$ might know that a certain ratio $\rho_k \beta_k \tau_p / \hat{\alpha}_{t,k}$ implies a certain probability of being the strongest UE. The bias term can then be made a random variable that depends on this ratio and makes the UE appoint itself the strongest user with the same probability (or a modified probability that, for example, prioritizes weaker UEs over stronger ones). Since the exact details depend strongly on the propagation environment and UE distribution, which are hard to compute and model exactly for practical setups, the design of probabilistic bias terms is mainly an engineering problem that is not studied further in this paper.
\end{remark}

\subsubsection{Allocation of Dedicated Payload Pilots}

The BS receives the repeated RA pilot transmissions in Step 3, which are followed by UL messages that, for example, can contain the unique identity number of the UE. The BS uses the $t$th pilot signal to estimate the channel to the UE (or UEs) that sent $\boldsymbol{\psi}_t$ in Step 3 and tries to decode the corresponding message. If the decoding is successful, the BS has identified one of the UEs in $\mathcal{S}_t$ and can admit it to the 
payload coherence blocks by allocating a pilot sequence (which typically is unique within the cell). This resource allocation decision is transmitted in the DL in Step 4, similarly to the precoded response in Step 2. The transmission can also contain other important information for the subsequent data transmission, such as timing advance. If the decoding fails, the SUCRe protocol has failed to resolve the collision. Note that if pilot $t$ was unused in Step 1 (i.e., $| \mathcal{S}_t |=0$), it will also be unused in Step 3 and hence the ``decoding'' will fail---there is no need for the BS to explicitly determine that $| \mathcal{S}_t |=0$.

The SUCRe protocol is repeated at a given interval. The UEs that were not admitted in Step 4 can be instructed when and how to transmit new RA pilots, for example, after a random waiting time. Alternatively, we can add additional steps on top of the SUCRe protocol to resolve the remaining collisions, by utilizing any conventional contention resolution method. For example, the UEs that collided in Step 3 can select new UL pilots at random and the risk of new collisions is vastly reduced since there are few remaining collisions.

\begin{remark}[Fairness] \label{remark:fairness}
If the UEs transmit at constant power, then the SUCRe protocol gives priority to UEs with the strong channel gains, which are typically in the cell center, while cell-edge UEs have weaker channel gains and are more likely to lose a contention. 
This short-term fairness issue is the price to pay for being able to resolve many collisions; the simulations in Section~\ref{section:numerical-results} show that around $0.9 \tau_p$ UEs out of the maximum $\tau_p$ are admitted in each RA block---even under high load. Since only a few UEs need to make new attempts, there will be fewer collisions in the long-run and also the weakest UEs will succeed after a few attempts (see Section~\ref{subsec:crowded-scenario} for numerical evidence). If more short-term fairness is desired (e.g., for more rapid handover), the decision rule can be changed towards this end; for example, by using probabilistic bias terms (see Remark~\ref{remark:bias-terms}) and/or uplink power control (see Remark~\ref{remark:power-control} below). 
\end{remark}

\section{Performance with Uncorrelated Rayleigh Fading Channels}
\label{sec:Rayleigh-fading}

In this section, we consider the special case of uncorrelated Rayleigh fading channels with 
\begin{equation} \label{eq:uncorr-Rayleigh}
\vect{h}_k \sim \mathcal{CN}(\vect{0},\beta_k \vect{I}_M ) 
\end{equation} 
for all UEs $k \in \mathcal{K}_0$. Furthermore, the inter-cell interference term is modeled as $\boldsymbol{\upsilon}_k \sim \mathcal{CN}(\vect{0},\Upsilon_k \vect{I}_{\tau_p})$ and is independent of the other signals. This model allows for a relatively tractable performance analysis for any value of $M$, in contrast to the analysis in Section~\ref{sec:proposed-protocol} that focused on $M$ for which the channel hardening and asymptotic favorable propagation properties are applicable (typically: $M > 50$). Recall that the received signal $z_k \in \mathbb{C}$ at UE~$k \in \mathcal{S}_t$ in Step~2 was given in \eqref{eq:received-user-k-scalar}.
The following lemma characterizes the distribution of this random variable.

\begin{lemma} \label{lemma:distribution_z_k} Consider uncorrelated Rayleigh fading channels.
For any UE~$k \in \mathcal{S}_t$ the received signal in \eqref{eq:received-user-k-scalar} can be expressed as $z_k = g_k + \nu_k$, where
\begin{align}
g_k &= \sqrt{ \frac{1}{2} \frac{\rho_k q \beta_k^2 \tau_p^2}{\alpha_t  + \sigma^2} } x, \quad x \sim \chi_{2M} \\
\nu_k &\sim \mathcal{CN} \left( 0, \,  \sigma^2 + \Upsilon_k +
q  \beta_k \tau_p - \frac{\rho_k q \beta_k^2 \tau_p^2}{\alpha_t  + \sigma^2}   \right)
\end{align}
are independent and $\chi_{n}$ denotes a chi-distribution with $n$ degrees of freedom.
\end{lemma}
\begin{IEEEproof}
The proof is given in Appendix \ref{app:collection-proofs}.
\end{IEEEproof}

By using the statistical properties in Lemma \ref{lemma:distribution_z_k}, we can compute the mean and variance of the normalized received DL signal $z_k/ \sqrt{M}$:
\begin{align} \label{eq:mean-zk/sqrtM}
& \mathbb{E}\left\{ \frac{z_k}{\sqrt{M}} \right\}  = \sqrt{ \frac{\rho_k q \beta_k^2 \tau_p^2}{\alpha_t  + \sigma^2} } \frac{\Gamma \left( M + \frac{1}{2} \right) }{ \sqrt{M} \Gamma \left( M \right) } , \\
&\mathbb{V} \left\{ \frac{z_k}{\sqrt{M}} \right\}    = \frac{\rho_k q \beta_k^2 \tau_p^2}{\alpha_t  + \sigma^2} \left( 1 - \left( \frac{\Gamma \left( M + \frac{1}{2} \right) }{\sqrt{M} \Gamma \left( M \right) } \right)^2  \right)  \notag \\ & + \frac{1}{M}\left(  \sigma^2 + \Upsilon_k +
q  \beta_k \tau_p - \frac{\rho_k q \beta_k^2 \tau_p^2}{\alpha_t  + \sigma^2}  \right). \label{eq:variance-zk/sqrtM}
\end{align}
By taking the limit  $M \rightarrow \infty$ and treating the Gamma function using Lemma~\ref{lemma:gamma-property} in Appendix~\ref{app:useful}, we obtain
\begin{align}
\mathbb{E}\left\{ \frac{z_k}{\sqrt{M}} \right\} & \rightarrow \sqrt{ \frac{\rho_k q \beta_k^2 \tau_p^2}{\alpha_t  + \sigma^2}  }, \quad \textrm{as} \quad M \rightarrow \infty, \label{eq:mean-value-z_k-asymptotics} \\
\mathbb{V} \left\{ \frac{z_k}{\sqrt{M}} \right\}  & \rightarrow 0, \quad \textrm{as} \quad M \rightarrow \infty.
\end{align}
The mean value approaches the limit in \eqref{eq:asymptotics-received-DL-channel} and the variance goes to zero, which confirms that Rayleigh fading channels offer asymptotic favorable propagation.

\subsection{Different Estimators of $\alpha_{t}$}

Instead of exploiting the channel hardening and asymptotic favorable propagation to estimate $\alpha_{t}$, as we did in  \eqref{eq:alpha-approx-est}, we can use the exact statistics from Lemma  \ref{lemma:distribution_z_k} to obtain the maximum likelihood (ML) estimate.

\begin{theorem} \label{theorem:ML-estimate-alpha} Consider uncorrelated Rayleigh fading channels.
The ML estimate of $\alpha_{t}$ from the observation $z_k = z_{k,\Re} + \jmath z_{k,\Im} $ (with $z_{k,\Re}, z_{k,\Im} \in \mathbb{R}$)  is
\begin{equation}
\hat{\alpha}_{t,k}^{\text{ML}} = \argmax{\alpha \geq \rho_k   \beta_k \tau_p} \quad f_1 \left( z_{k,\Re} | \alpha \right) f_2 \left( z_{k,\Im} | \alpha \right) 
\end{equation}
for the conditional probability density functions (PDFs)
\begin{align}  \label{eq:f1-pdf}
&f_1 \left( z_{k,\Re} | \alpha \right) =  \frac{ e^{- \frac{(z_{k,\Re})^2}{\lambda_2}  \left( 1 - \frac{\lambda_1}{\lambda_1 + \lambda_2}   \right)} }{\Gamma(M) \lambda_1^M \sqrt{\pi \lambda_2 }} \sum_{n=0}^{2M-1} {2M-1 \choose n}   \notag \\
& \times  \frac{\left( \Gamma \left( \frac{n+1}{2} \right) + c_n(z_{k,\Re}) \gamma \left( \frac{n+1}{2} , \frac{(z_{k,\Re})^2}{\lambda_2} \frac{\lambda_1}{\lambda_1+\lambda_2}  \right)  \right) }{   \left( \frac{z_{k,\Re}}{\lambda_2} \right)^{n+1-2M}  \left(  \frac{1}{\lambda_1} + \frac{1}{\lambda_2} \right)^{2M-\frac{n+1}{2}} }  
\\
&f_2 \left( z_{k,\Im} | \alpha \right) = \frac{1}{\sqrt{\pi \lambda_2 }} e^{- \frac{( z_{k,\Im} )^2 }{\lambda_2} }, \label{eq:f2-pdf}
\end{align}
where $\gamma(\cdot,\cdot)$ is the lower incomplete gamma function\footnote{\label{footnote:incomplete-gamma}For any positive integer $m$, 
the lower incomplete gamma function can be computed as $\gamma(m,x) = \Gamma(m) - \Gamma(m) e^{-x} \sum_{k=0}^{m-1} x^k/\Gamma(k+1) $.}, 
\begin{equation} \label{eq:cn-definition}
c_n(z) = \begin{cases}
(-1)^n & z \geq 0, \\
-1 & z<0,
\end{cases}
\end{equation}
 and the coefficients $\lambda_1$ and $\lambda_2$ depend on $\alpha$ as
\begin{align} \label{eq:lambda1}
\lambda_1 &= \frac{\rho_k q \beta_k^2 \tau_p^2}{\alpha  + \sigma^2}    \\
\lambda_2 &= \sigma^2 + \Upsilon_k + q  \beta_k \tau_p  - \lambda_1. \label{eq:lambda2}
\end{align}
\end{theorem}
\begin{IEEEproof}
The proof is given in Appendix \ref{app:collection-proofs}.
\end{IEEEproof}

The ML estimate $\hat{\alpha}_{t,k}^{\text{ML}}$ of $\alpha_{t}$ can be computed numerically using Theorem \ref{theorem:ML-estimate-alpha}.\footnote{The conditional PDF $f_1 \left( z_{k,\Re} | \alpha \right) $ contains several terms that grow rapidly with $M$, while their ratios remain small. Hence, a careful implementation of the PDF is needed for numerical stability. The simulations in this paper were implemented successfully by taking the logarithm of each term in the summation.} 
An approximate estimate in closed-form can also be obtained from \eqref{eq:mean-zk/sqrtM} by utilizing the fact that 
\begin{equation}
\Re( z_{k} ) \approx  \mathbb{E}\left\{ z_k \right\} = \sqrt{ \frac{\rho_k q \beta_k^2 \tau_p^2}{\alpha_t  + \sigma^2} } \frac{\Gamma \left( M + \frac{1}{2} \right) }{ \Gamma \left( M \right) }
\end{equation}
when $M$ is large. By solving the equation $\Re( z_{k} ) =  \mathbb{E}\left\{ z_k \right\}$ for $\alpha_{t}$ we obtain the estimator
\begin{equation} \label{eq:alpha-approx-est2}
\hat{\alpha}_{t,k}^{\text{approx2}} = \max \bigg( \! \bigg( \frac{\Gamma \left( M \!+\! \frac{1}{2} \right) }{ \Gamma \left( M \right) } \bigg)^2
\frac{ q \rho_k   \beta_k^2 \tau_p^2  }{ \left( \Re ( z_k ) \right)^2 } - \sigma^2, \, \rho_k   \beta_k \tau_p  \bigg)
\end{equation}
since UE~$k$ knows that $\alpha_t \geq \rho_k  \beta_k \tau_p$.
This estimator is slightly different from the one obtained in \eqref{eq:alpha-approx-est}, but they are asymptotically equivalent as $M \rightarrow \infty$ due to \eqref{eq:gamma-property} and both are asymptotically error free.

\begin{figure} \vspace{-3mm}
        \centering
        \begin{subfigure}[b]{.95\columnwidth} \centering 
                \includegraphics[width=\columnwidth]{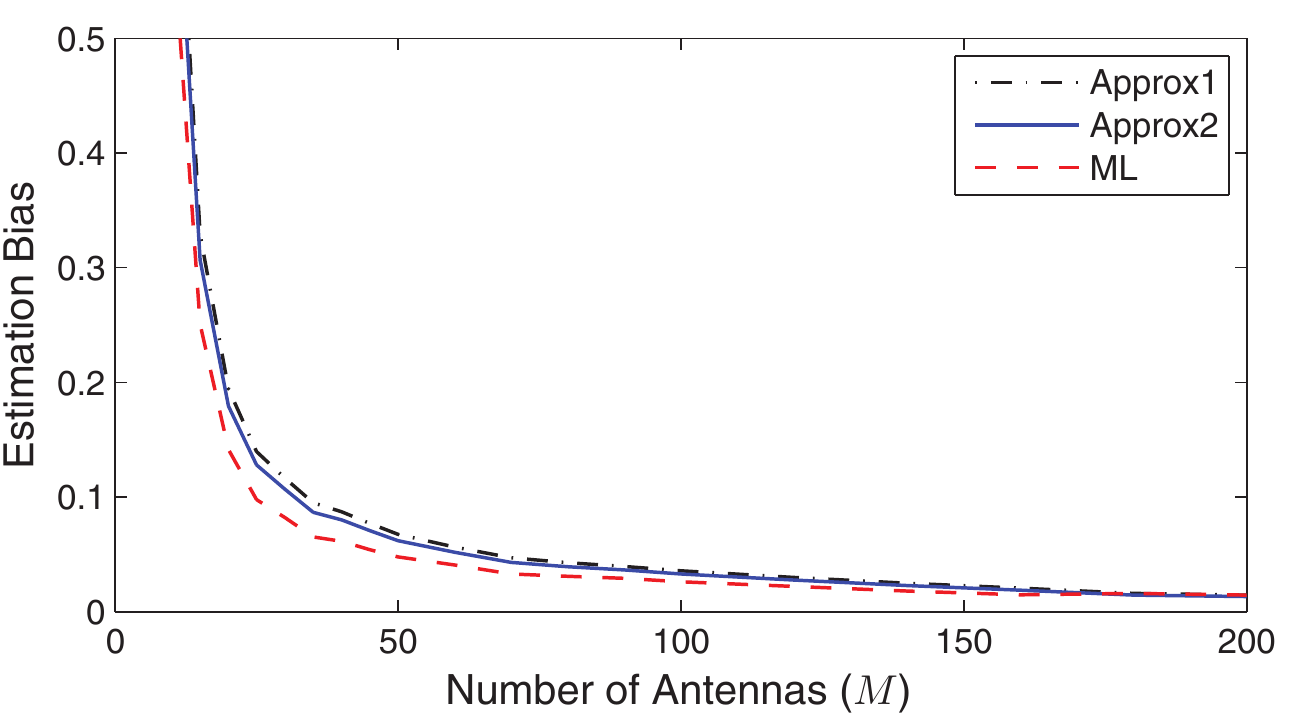} \vspace{-5mm}
                \caption{Normalized bias}
                \label{figure5a}
        \end{subfigure} \hspace{3mm}
        \begin{subfigure}[b]{.95\columnwidth} \centering
                \includegraphics[width=\columnwidth]{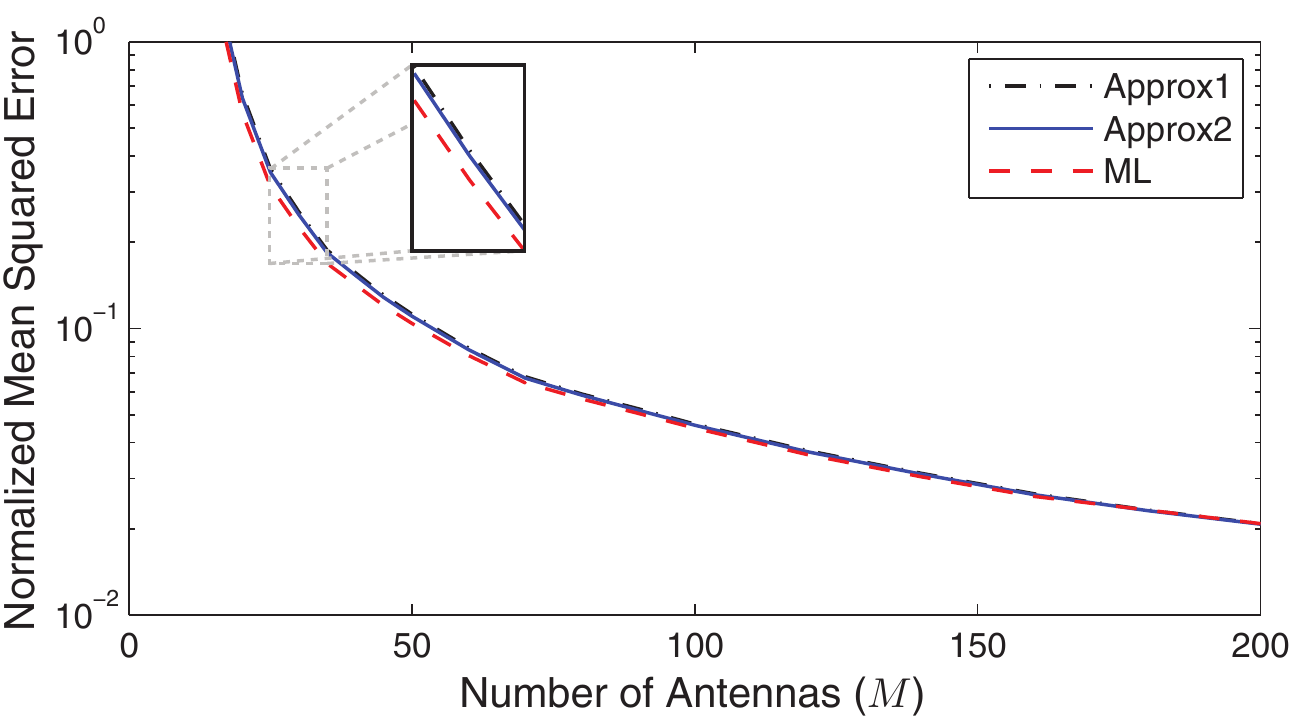} \vspace{-5mm}
                \caption{Normalized mean squared error}
                \label{figure5b} 
        \end{subfigure} 
        \caption{Comparison of three estimators of the signal gain $\alpha_t$: $\hat{\alpha}_{t,k}^{\text{ML}}$, $\hat{\alpha}_{t,k}^{\text{approx1}}$, and $\hat{\alpha}_{t,k}^{\text{approx2}}$. The true value is $\alpha_t = 20$, while the pilot SNR at UE~$i$ is 10\,dB.}\label{fig:estimatorcomparison}
\end{figure}

We have obtained three different estimators of $\alpha_{t}$: $\hat{\alpha}_{t,k}^{\text{ML}}$, $\hat{\alpha}_{t,k}^{\text{approx1}}$, and $\hat{\alpha}_{t,k}^{\text{approx2}}$. Which one is preferred in practice? The performance of these estimators is compared in Fig.~\ref{fig:estimatorcomparison}, where  $\alpha_t = 20$ and UE~$i$ estimates $\alpha_{t}$ while having $q= \rho_i \beta_i=\sigma^2=1$ and $\tau_p=10$. This corresponds to an SNR of 0\,dB, while the effective pilot SNR is 10\,dB. Fig.~\ref{fig:estimatorcomparison}(a) shows the normalized bias $(\mathbb{E}\{\hat{\alpha}_{t,k} \}-\alpha_t)/\alpha_t$ and Fig.~\ref{fig:estimatorcomparison}(b) shows the normalized mean-squared error (NMSE) $\mathbb{E}\{| \hat{\alpha}_{t,k} -\alpha_t |^2 \}/\alpha_t$.
All three estimators perform badly for $M<25$, but become asymptotically unbiased as $M$ increases and achieve NMSEs below $10^{-1}$ for $M\geq 50$. The SUCRe protocol is thus particularly useful in Massive MIMO. All estimators have a tendency to overestimate $\alpha_t$, as seen from the positive bias. The ML estimator $\hat{\alpha}_{t,k}^{\text{ML}}$ provides the smallest NMSEs, as expected, while $\hat{\alpha}_{t,k}^{\text{approx2}}$ is better than $\hat{\alpha}_{t,k}^{\text{approx1}}$. However, the differences are tiny and thus we appoint $\hat{\alpha}_{t,k}^{\text{approx2}}$ as the preferred choice since the ML estimator is computationally involved.

\subsection{Probability of Pilot Repetition}

The core of the SUCRe protocol is that only one UE should transmit pilot $t$ in Step 3. The probability that a particular UE decides to repeat its pilot transmission is computed as follows.

\begin{theorem} \label{theorem:retransmission-probability}
If UE~$k \in \mathcal{S}_t$ estimates $\alpha_t$ as in \eqref{eq:alpha-approx-est2} and applies the proposed decision rule with\footnote{For all such bias terms, the estimate $\hat{\alpha}_{t,k} = \rho_k   \beta_k \tau_p$ leads to a decision to repeat the pilot, which makes perfect sense since the UE believes that it is the only one that transmitted the pilot.} $\epsilon_k < \rho_k   \beta_k \tau_p/2$, then the probability of repeating the pilot in Step 3 is
\begin{equation} \label{eq:probability-retransmit}
\Pr \{ \mathcal{R}_{k} \} = 1 - \Pr \{ \Re ( z_k )   \leq \sqrt{\zeta_k} \}  + \Pr \{ \Re ( z_k )   \leq - \sqrt{\zeta_k} \} 
\end{equation}
where
\begin{equation} \label{eq:zeta-parameter}
\zeta_k = \bigg( \frac{\Gamma \left( M \!+\! \frac{1}{2} \right) }{ \Gamma \left( M \right) } \bigg)^2 \frac{q \rho_k   \beta_k^2 \tau_p^2 }{\sigma^2+2( \rho_k \beta_k \tau_p - \epsilon_k) }
\end{equation}
and
\begin{align}  
& \Pr \{ \Re ( z_k )  \leq b \} = Q \left(-\frac{b \sqrt{2}}{\sqrt{\lambda_2}} \right) -  \sum_{k=0}^{M-1} \frac{ e^{- \frac{b^2}{\lambda_2}  \left( 1 - \frac{\lambda_1}{\lambda_1 + \lambda_2}   \right)} }{\Gamma(k+1) \lambda_1^k \sqrt{\pi \lambda_2 }}  \notag \\
& \times \sum_{n=0}^{2k} {2k \choose n}     \frac{\left( \Gamma \left( \frac{n+1}{2} \right) + c_n(b) \gamma \left( \frac{n+1}{2} , \frac{b^2}{\lambda_2} \frac{\lambda_1}{\lambda_1+\lambda_2}  \right)  \right) }{  2 \left( \frac{b}{\lambda_2} \right)^{n-2k}  \left(  \frac{1}{\lambda_1} + \frac{1}{\lambda_2} \right)^{2k+\frac{1}{2}-\frac{n}{2}} } .\label{eq:cdf-zre}
\end{align}
In these expressions, $Q(x)=\frac{1}{\sqrt{2 \pi}} \int_{x}^{\infty} e^{-t^2/2}dt$ is the $Q$-function, $c_n(\cdot)$ is defined in
\eqref{eq:cn-definition}, and $\lambda_1$ and $\lambda_2$ are given in \eqref{eq:lambda1}--\eqref{eq:lambda2} with $\alpha= \alpha_t$.
\end{theorem}
\begin{IEEEproof}
The proof is given in Appendix \ref{app:collection-proofs}.
\end{IEEEproof}

This theorem provides the CDF of $\Re ( z_k )$ and uses it to compute the exact probability $\Pr \{ \mathcal{R}_{k} \}$ that UE~$k$ repeats the pilot in Step 3. Note that $\Pr \{ \Re ( z_k )   \leq - \sqrt{\zeta_k} \}>0$ in general, since there is a (tiny) probability that the noise and inter-cell interference will make $\Re ( z_k )$ negative. Since the expressions in Theorem \ref{theorem:retransmission-probability} are fairly complicated, we also study the asymptotic behavior as $M \rightarrow \infty$.

\begin{corollary} \label{cor:asymptotic-CDFs}
For large $M$, the complementary CDF $\Pr \{ \Re ( z_k )  > b \} $ in \eqref{eq:cdf-zre} converges to
\begin{equation} \label{eq:asympotic-CDF}
Q \!\left( \frac{b-\frac{\Gamma \left( M \!+\! \frac{1}{2} \right) }{ \Gamma \left( M \right) }  \sqrt{\lambda_1} }{  \sqrt{ \lambda_1^2 \Big( M- \Big(\frac{\Gamma \left( M \!+\! \frac{1}{2} \right) }{ \Gamma \left( M \right) }  \Big)^2 \Big)   + \sigma^2 + \Upsilon_k + q  \beta_k \tau_p - \lambda_1 }  }  \right)
\end{equation}
where the $Q$-function was defined in Theorem \ref{theorem:retransmission-probability}.
In particular, for $\epsilon_k < \rho_k   \beta_k \tau_p/2$, the probability becomes
\begin{equation} \label{eq:asymptotic-retransmission-prob}
\lim_{M \rightarrow \infty} \Pr \{ \mathcal{R}_{k} \} = \begin{cases}
0, & \rho_k \beta_k \tau_p < \alpha_t/2 + \epsilon_k, \\ 
1/2, &\rho_k \beta_k \tau_p = \alpha_t/2 + \epsilon_k, \\
1, & \rho_k \beta_k \tau_p > \alpha_t/2 + \epsilon_k. 
\end{cases}
\end{equation}
\end{corollary}
\begin{IEEEproof}
The proof is given in Appendix \ref{app:collection-proofs}.
\end{IEEEproof}

This corollary confirms that if one UE has a signal gain that is larger than the sum of all the others' signal gains (plus inter-cell interference), then this UE will repeat its pilot in Step 3 with probability one in the regime of large number of antennas.
The bias parameter $\epsilon_k$ can be used to tune this condition; for example, at most one UE will transmit the pilot in Step 3 when $\epsilon_k>0$ for all $k$.
At finite $M$ there is always a risk for collisions and the pilot repetition decisions are correlated between the UEs, since they are based on partially the same sources of randomness (i.e., channel realizations and inter-cell interference). Since the expression for $\Pr \{ \mathcal{R}_{k} \} $ in Theorem \ref{theorem:retransmission-probability} is complicated, an exact computation of the probability to resolve collisions appears intractable.
We will thus study this numerically instead.

\begin{figure} \vspace{-3mm}
        \centering
        \begin{subfigure}[t]{.95\columnwidth} \centering 
                \includegraphics[width=\columnwidth]{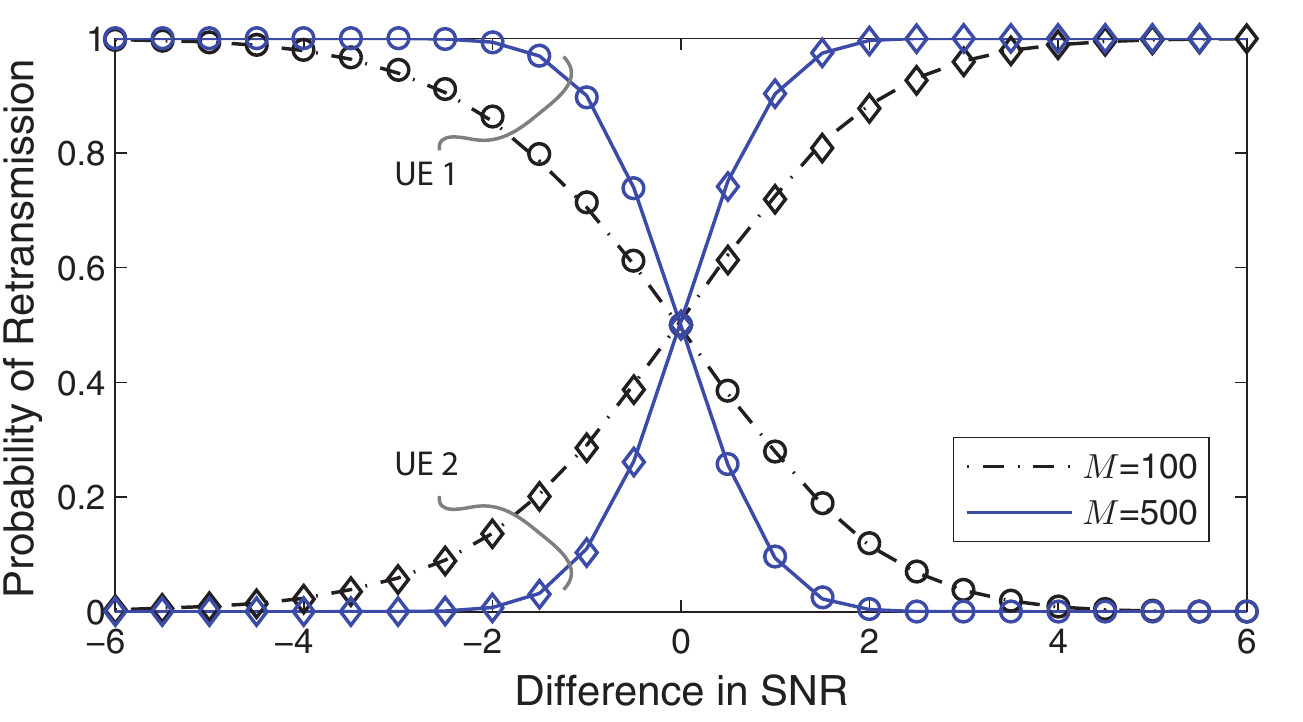} \vspace{-5mm}
                \caption{Probability to repeat the pilot transmission in a two-UE collision}
                \label{figure6a}
        \end{subfigure}  \hspace{3mm}
        \begin{subfigure}[t]{.95\columnwidth} \centering
                \includegraphics[width=\columnwidth]{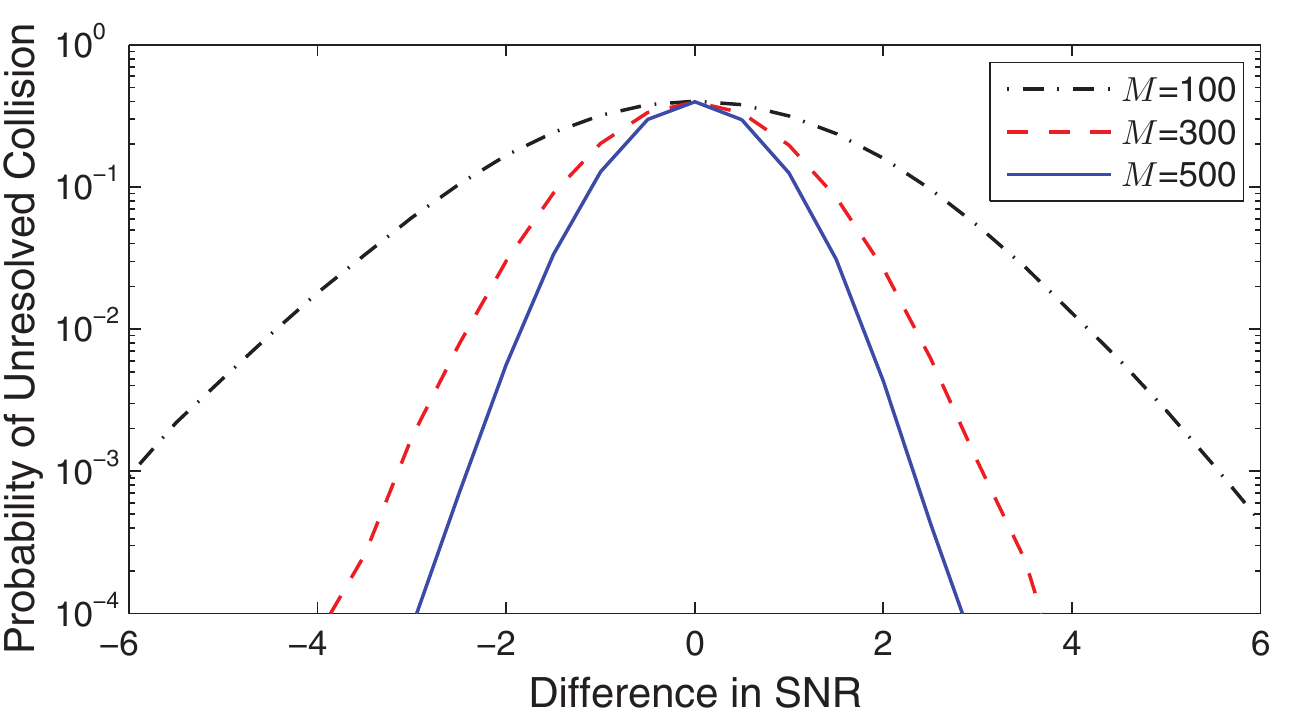} \vspace{-5mm}
                \caption{Probability of having an unresolved two-UE collision}
                \label{figure6b} 
        \end{subfigure} 
        \caption{Two-UE collisions are studied where UE 1 has $\rho_1 \beta_1 \tau_p=10$\,dB and UE 2 has $\rho_2 \beta_2 \tau_p$ between $4$\,dB and $16$\,dB. The SNR difference is $\rho_2 \beta_2 \tau_p-\rho_1 \beta_1 \tau_p$.  (a) shows the probabilities that each of the UEs repeat the pilot in Step 3; (b) shows the probability of having an unresolved collision.}\label{fig:basic-retransmission}
\end{figure}

\subsection{Example: Resolving a Two-UE Collision}

Let us consider a case where two UEs collide: $\mathcal{S}_t=\{1,2\}$. The first UE has the fixed pilot SNR $\mathrm{SNR}_1 = \rho_1 \beta_1 \tau_p = q \beta_1 \tau_p = 10$\,dB, while the corresponding $\mathrm{SNR}_2$ of the second UE is varied between $4$\,dB and $16$\,dB (with normalized noise variance $\sigma^2=1$). Fig.~\ref{fig:basic-retransmission}(a) shows the probability that the UEs repeat their pilot transmissions in Step 3, assuming $\epsilon_k = 0$ and either $M=100$ or $M=500$ BS antennas.  The horizontal axis shows the SNR difference $\mathrm{SNR}_2 - \mathrm{SNR}_1$ between the UEs, which is between $-6$\,dB and $+6$\,dB. The curves were generated by Monte-Carlo simulations, while the markers are computed using the closed-form expression in Theorem \ref{theorem:retransmission-probability}. The minor discrepancies are due to the finite number of Monte Carlo realizations. If there is an SNR difference of at least 3\,dB, then the UE with the largest SNR is likely the only one to repeat the pilot. In contrast, both UEs repeat the pilot with equal probability when they have identical SNR. The transition between these cases is sharper the more antennas are used, which is in line with Corollary  \ref{cor:asymptotic-CDFs}.

The probability of having an unresolved two-UE collision, $1-P_{2,\textrm{resolved}} $ with $P_{2,\textrm{resolved}}$ defined in \eqref{eq:prob-resolved-S2}, is shown in Fig.~\ref{fig:basic-retransmission}(b) for $\epsilon_k = 0$ and $M \in \{100, \, 300, \, 500\}$. The proposed SUCRe protocol resolves almost all collisions when $\mathrm{SNR}_1$ and $\mathrm{SNR}_2$ are sufficiently different (e.g., more than $90\%$ of the two-UE collisions when there is a 3\,dB difference in SNR). The probability of having an unresolved collision drops rapidly when adding more antennas, except in the special case of $\mathrm{SNR}_1 = \mathrm{SNR}_2$ where it is stable at around $40\%$. If the decisions would have been independent between the UEs, then $50\%$ of the collisions would be unresolved in this special case. Hence, the errors in the estimates $\hat{\alpha}_{t,1}^{\text{approx2}} $ and $\hat{\alpha}_{t,2}^{\text{approx2}} $ are correlated; when one UE experiences a small-scale channel realization much stronger than the average it will underestimate $\alpha_{t}$, while the other UE is likely to overestimate $\alpha_{t}$ since it will believe that the other UE has a stronger average channel gain than it actually has.

This example shows that SNR differences are desirable when using the SUCRe protocol, which means that we should embrace rather than fully combat the pathloss variations that appear between UEs in cellular networks. We elaborate on this in the following remark.

\begin{remark}[Uplink pilot power control] \label{remark:power-control}
Power control is used in conventional RA, such as LTE, to give the UEs equal conditions (i.e., the same signal gain $\rho_k \beta_k$). UEs at the cell edge transmit at full power, while cell-center UEs reduce their power. The SUCRe protocol prefers the opposite type of power control; each UE transmits at full power and the pathloss variations are exploited to resolve potential RA collisions. If the pathloss difference are larger than the dynamic range of the BS receiver, some fractional power control is needed to partially reduce the pathloss differences. In many practical cases it is very likely that the colliding UEs have pathloss differences (in terms of $\beta_k$) of at least 3\,dB. In situations where this is not the case, the RA pilot powers $\rho_k$ can be randomized around a nominal value to create a similar type of randomness where the colliding UEs are unlikely to have the same SNR. We will illustrate this numerically in the next section. Randomized power control can also increase the fairness (see Remark~\ref{remark:fairness}) by making the probability of being the strongest UE less dependent on $\beta_k$, which can be important in systems where random access is used for decentralized handover between cells.
Ideal fairness can, in principle, be achieved by first applying conventional power control that makes $\rho_k \beta_k$ equal for all UEs and then generate random power variations around this nominal  value.
\end{remark}

\section{Numerical Results}
\label{section:numerical-results}

In this section, we show numerically how the SUCRe protocol performs in cellular networks. The simulation results can be reproduced using the Matlab code that is available at \url{https://github.com/emilbjornson/sucre-protocol}. We consider the center cell of the hexagonal network depicted in Fig.~\ref{figure:block-structure-multi-cell} and take the activities in the six neighboring cells into account. The radius of each hexagon is 250 m, and the UEs are uniformly distributed in each cell  at locations further than 25 m from the BS. The estimate $\hat{\alpha}_{t,k}^{\text{approx2}}$ of $\alpha_{t}$ is used in all the simulations.

\subsection{Channel Propagation Models}

We will compare three different channel models. The first one is uncorrelated Rayleigh fading, where $\vect{h}_k $ is distributed as in
\eqref{eq:uncorr-Rayleigh} for $k=1,\ldots,K_0$. The second one is correlated Rayleigh fading with $\vect{h}_k \sim \mathcal{CN}(\vect{0},\beta_k \vect{R}_k )$, where we consider a ULA at the BS modeled by the exponential correlation model with the correlation $r=0.7$ between adjacent antennas \cite{Loyka2001a}:
\begin{equation}
\left[ \vect{R}_k \right]_{i,j} = r^{- | j-i |} e^{\jmath \theta_k (j-i)}
\end{equation}
where $\theta_k$ is the angle between UE~$k$ and BS~$0$. This represents a non-line-of-sight scenario with spatial correlation, meaning that the channel is statistically stronger in some spatial directions (determined by $\theta_k$) than other directions.

The third model describes pure line-of-sight (LoS) propagation when the BS is equipped with a ULA with half-wavelength antenna spacing:\footnote{All elements in \eqref{eq:LoS-vector-channel} should also be multiplied with a common phase-shift, but we neglect it here since it has no impact on the performance.}
\begin{equation} \label{eq:LoS-vector-channel}
\vect{h}_{k} = \sqrt{\beta_{k}} \left[ 1 \,\,\, e^{-  \jmath \pi  \sin(\phi_{k}) } \,\, \ldots \,\,  e^{-  \jmath  \pi (M-1) \sin(\phi_{k}) }  \right]^{\Ttran}.
\end{equation}
Note that this channel vector is deterministic, in contrast to the previous two models.

The pathloss is modeled based on the urban micro scenario in \cite{LTE2015}. The Rayleigh fading cases have pathloss exponent 3.8 and shadow fading with log-normal distribution and standard deviation 10\,dB, while the LoS case has pathloss exponent 2.5 and log-normal variations with standard deviation 4\,dB. When a UE in a corner point of the cell transmits at full power the median of the SNR, $\rho_k \beta_k / \sigma^2$, is 0\,dB in the non-LoS case and 33\,dB in the LoS case. The BS and UEs transmit at the same constant power (i.e., $\rho_k = q$) giving the same SNR in both directions.

We will consider both cases when the adjacent cells are silent during the RA protocol and when they perform regular data transmission. In the latter case, we assume that there are ten active UEs in each of the neighboring cells and the propagation channels are modeled as uncorrelated Rayleigh fading (using the same power levels and pathloss models as above).\footnote{We make sure that each BS has a hexagonal coverage area by only considering shadow fading realizations where each UE gets the highest signal gain from its serving BS.} The average UL interference $\bar{\omega} =\mathbb{E}\{ \| \vect{W} \frac{\boldsymbol{\psi}_t^*}{\| \boldsymbol{\psi}_t \|} \|^2 /M \}$, where the expectation is computed with respect to user locations and shadow fading realizations, is assumed to be known at the UE (it is the same for all UEs) and is subtracted from $\hat{\alpha}_{t,k}$ by setting $\epsilon_k = - \bar{\omega}/2$.

\begin{figure}[t!]
\begin{center}
\includegraphics[width=\columnwidth]{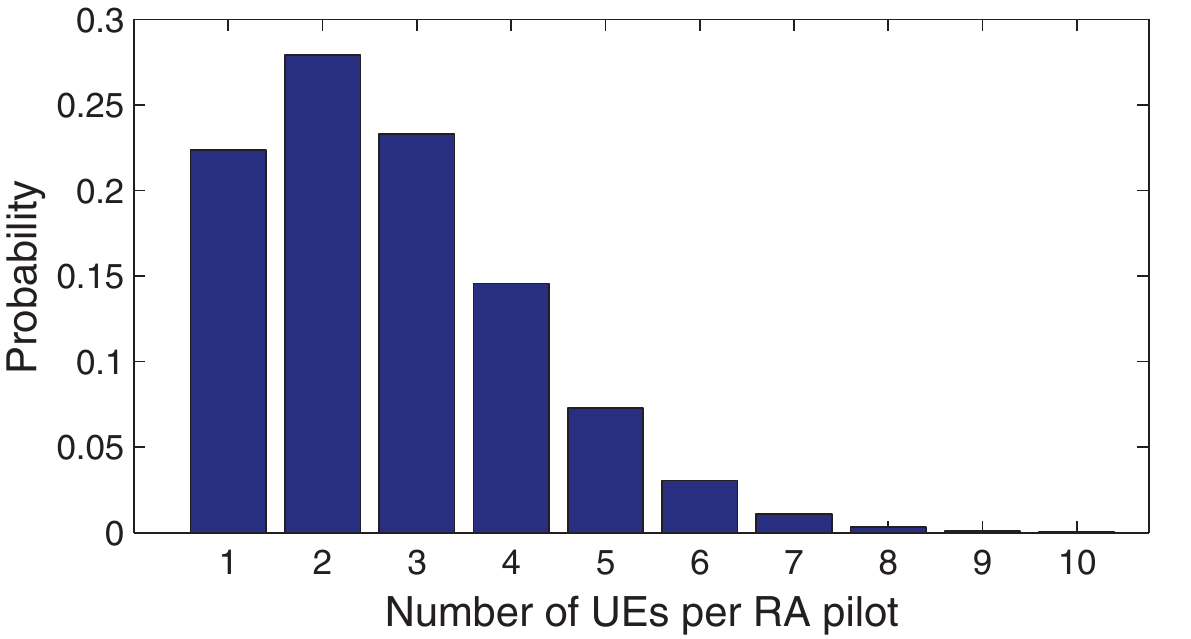}
\end{center} \vskip-3mm
\caption{Example of a distribution of the number of UEs that selects each RA pilot. It is used in Figs.~\ref{fig:cellular-simulation} and \ref{fig:bias-simulation}.} \label{figure:usersperpilot} \vskip-3mm
\end{figure}

\begin{figure} \vspace{-3mm}
        \centering
        \begin{subfigure}[b]{.95\columnwidth} \centering 
                \includegraphics[width=\columnwidth]{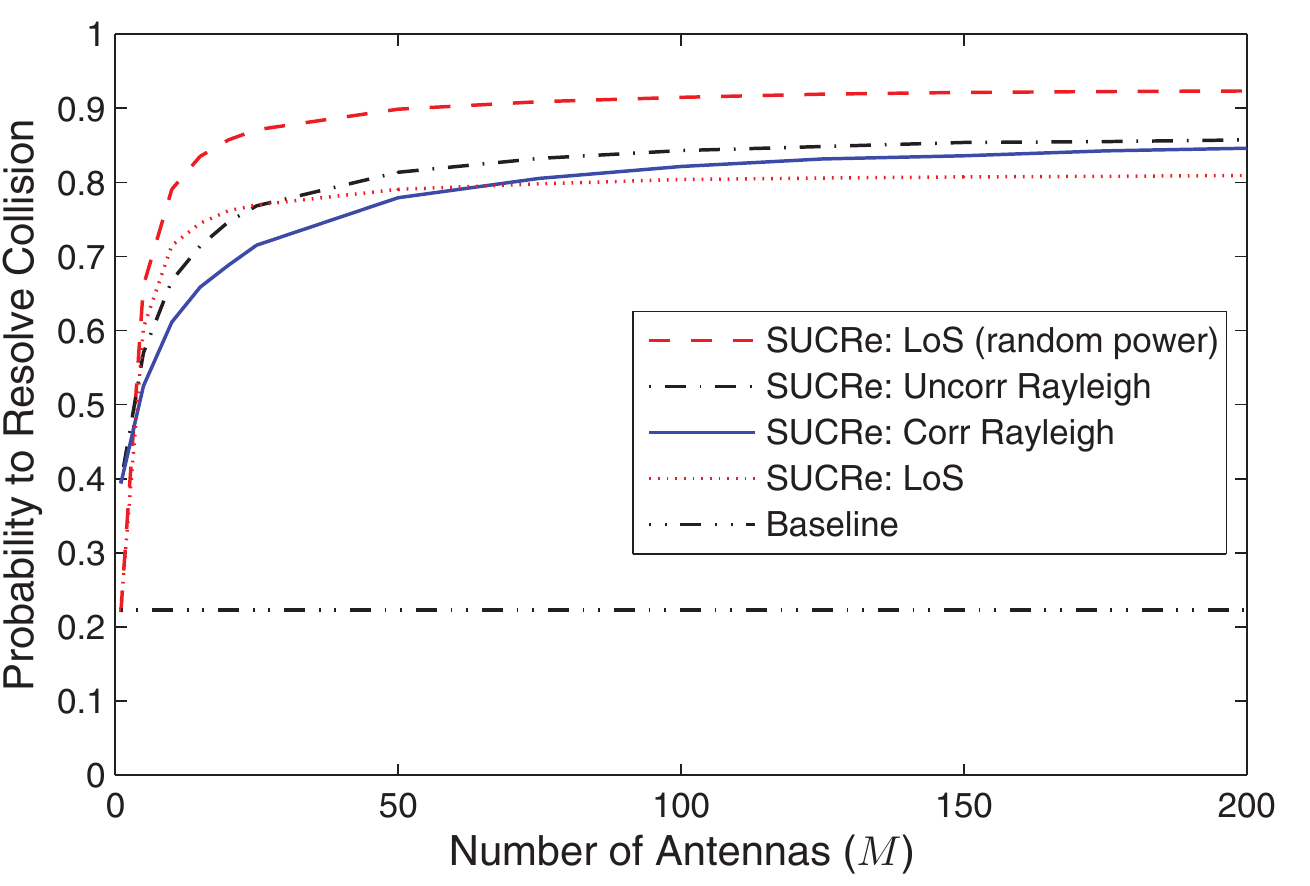} \vspace{-5mm}
                \caption{With interference from adjacent cells} 
                \label{figure8a}
        \end{subfigure}  \hspace{3mm}
        \begin{subfigure}[b]{.95\columnwidth} \centering
                \includegraphics[width=\columnwidth]{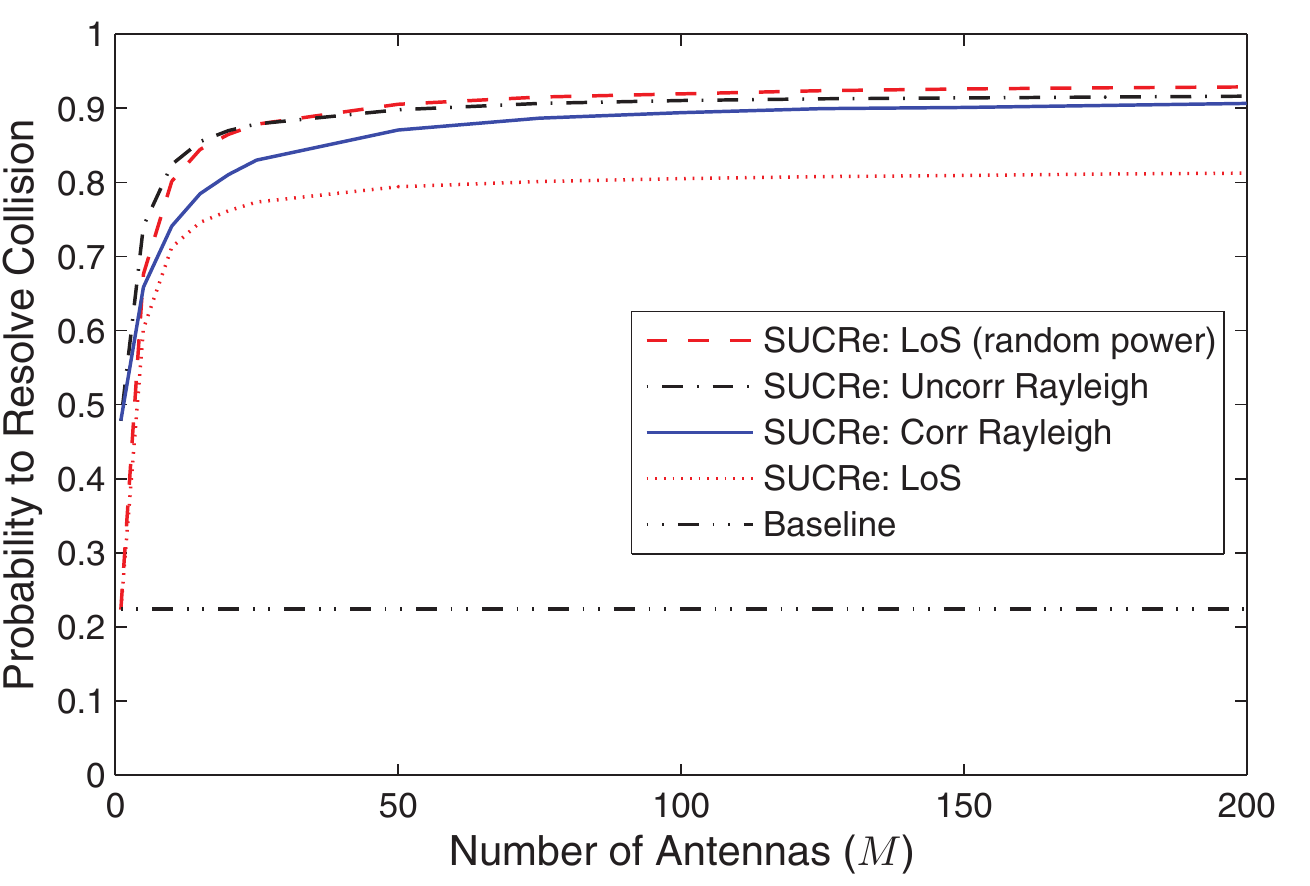} \vspace{-5mm}
                \caption{Without interference from adjacent cells}  
                \label{figure8b} 
        \end{subfigure} 
        \caption{Probability of resolving collisions, as a function of the number of BS antennas, in a highly loaded cellular network with or without inter-cell interference.}\label{fig:cellular-simulation}  \vspace{-3mm}
\end{figure}

\subsection{Probability to Resolve Collisions}

We will now illustrate that the SUCRe protocol is capable of resolving collisions, even when the system is overloaded. We consider a scenario with $\tau_p=10$ and $K_0 = 5000$ inactive UEs in the cell, where each UE accesses the network with a $0.5 \%$ probability in a given RA block. 
The number of UEs, $| \mathcal{S}_t |$, is distributed as illustrated in Fig.~\ref{figure:usersperpilot}, which is obtained from the binomial distribution in \eqref{eq:binomial-dist-N} by conditioning on that $| \mathcal{S}_t | \geq 1$. This is an overloaded scenario in the sense that, on average, 2.5 UEs select each pilot, leading to collisions at more than 75\% of the pilots. Even collisions with 5 or 6 UEs occur frequently.

The probability to resolve collisions is defined as
\begin{align} \notag
&P_{\textrm{resolved}} = \mathbb{E} \left\{ P_{|\mathcal{S}_t|,\textrm{resolved}}  \big|  |\mathcal{S}_t| \geq 1 \right\}  \\
&= \sum_{N =1}^{K_0} \frac{P_{N,\textrm{resolved}} \,  {K_0 \choose N}
\left( \frac{P_a}{\tau_p} \right)^N \! \left(1-\frac{P_a}{\tau_p} \right)^{K_0-N} }{ 1 - \left(1-\frac{P_a}{\tau_p} \right)^{K_0}  },
\end{align}
where $P_{N,\textrm{resolved}}$ was defined in  \eqref{eq:prob-resolved} and the expectation is with respect to $|\mathcal{S}_t|$.
Fig.~\ref{fig:cellular-simulation} shows this probability to resolve collisions as a function of the number of BS antennas, for the aforementioned three channel models. Fig.~\ref{fig:cellular-simulation}(a) considers inter-cell interference, while Fig.~\ref{fig:cellular-simulation}(b) neglects the interference (i.e., the adjacent cells are silent in the RA block).

The first observation from Fig.~\ref{fig:cellular-simulation} is that the SUCRe protocol relies on the channel hardening and favorable propagation of Massive MIMO channels; $P_{\textrm{resolved}} $ is $20$-$40\%$ at $M=1$, but increases steeply to $75$-$90\%$ when having $M=50$ antennas. The probability to resolve collisions continues to increase for $M \geq 50$, but at a slower pace. Uncorrelated Rayleigh fading gives better results than correlated Rayleigh fading, but the difference is small when $M$ is large. The LoS model has slightly worse performance, because the pathloss differences are lower which makes it harder to appoint a strongest UE. However, since the cell-edge SNR is higher in the LoS case we can afford to create SNR differences by randomizing the UL pilot powers (as discussed in Remark \ref{remark:power-control}). Fig.~\ref{fig:cellular-simulation} also shows the LoS case when each UE reduces its pilot power with a random number between $0$\,dB and $-30$\,dB, uniformly distributed in dB-scale. This case gives the highest performance among all the cases. Hence, the SUCRe protocol is well suited for both LoS and non-LoS channels.

As a baseline, we also consider a conventional protocol where pilot collisions are only handled by retransmission in later RA blocks. 
Specifically, using Fig.~\ref{figure:PRACH}, all UEs deterministically repeat the pilots in Step 3 of the access protocol and proceed to the process of centralized contention resolution. 
Fig.~\ref{fig:cellular-simulation} shows the probability to resolve collisions and we notice that the SUCRe protocol is able to admit roughly four times as many UEs per RA pilot than the baseline scheme. We also notice that inter-cell interference only causes a minor degradation in performance, except in the LoS case where performance is almost unaffected. The full-power LoS case performs similarly to the other channel models in this case.

\begin{figure} \vspace{-3mm}
        \centering
        \begin{subfigure}[b]{.95\columnwidth} \centering 
                \includegraphics[width=\columnwidth]{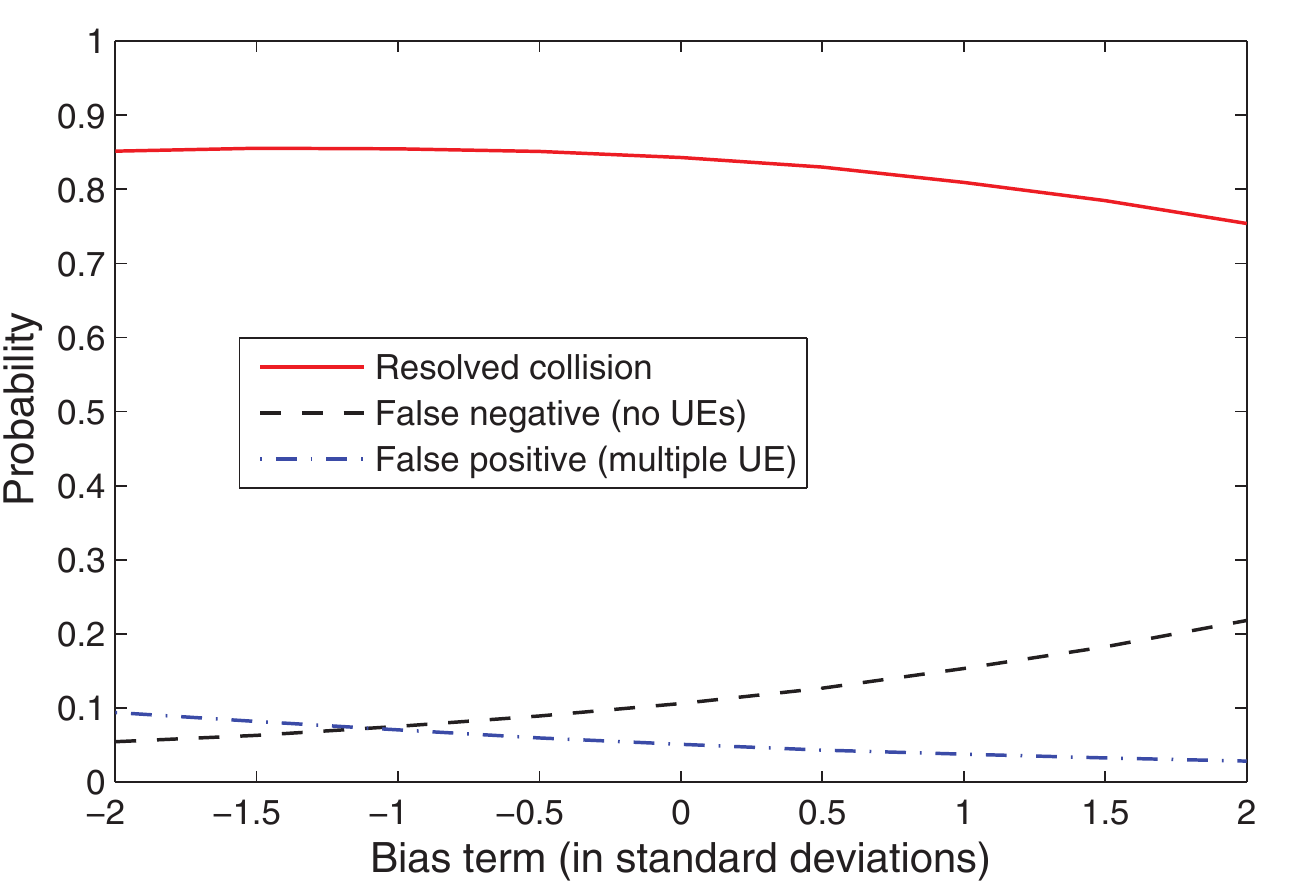} \vspace{-5mm}
                \caption{With interference from adjacent cells}
                \label{figure9a}
        \end{subfigure}  \hspace{2mm}
        \begin{subfigure}[b]{.95\columnwidth} \centering
                \includegraphics[width=\columnwidth]{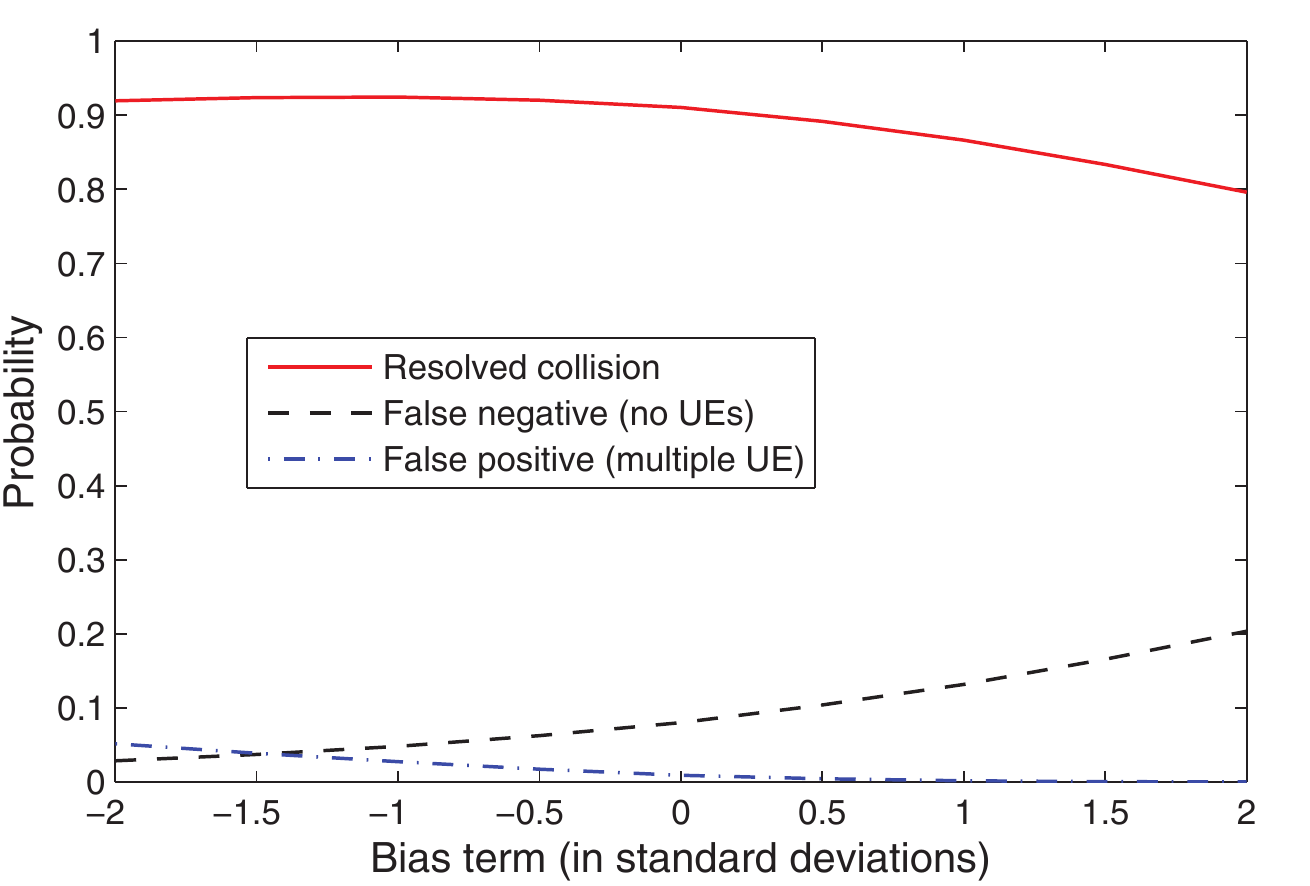} \vspace{-5mm}
                \caption{Without interference from adjacent cells}
                \label{figure9b} 
        \end{subfigure} 
        \caption{Probability of resolving collisions, false negatives, and false positives for different bias terms in the decision rule, for uncorrelated Rayleigh fading channels.}\label{fig:bias-simulation} \vspace{-3mm}
\end{figure}

\subsection{Tuning Probabilities using Bias Term}

Next, we consider the same scenario but focus on uncorrelated Rayleigh fading channels and study the probabilities of resolving collisions, false negatives, and false positives. We will demonstrate how the bias term $\epsilon_k$ can be utilized to tune the decisions by setting $\epsilon_k = \delta \beta_k/ \sqrt{M} - \bar{\omega}/2$, which corresponds to adding $\delta$ standard deviations of $\| \vect{h}_k \|^2 / M$ (around its mean value $\beta_k$). Fig.~\ref{fig:bias-simulation} shows the probabilities as a function of $\delta$, for $M=100$ and with or without inter-cell interference.

By subtracting one or two standard deviations from $ \hat{\alpha}_{t,k} /2$ in the decision rule, we can encourage UE~$k$ to appoint itself the contention winner. This leads to higher probability of resolving collisions, at the cost of more false positives where multiple UEs repeat their pilot transmissions in Step~3. In contrast, by adding one or two standard deviations to $\hat{\alpha}_{t,k} /2$ in the decision rule, we can discourage UE~$k$ from appointing itself the contention winner and thereby push the probability of false positives towards zero---at the cost of resolving fewer collisions and having more false negatives where no UEs repeat their pilots in Step 3.

The probability of resolving a collision is naturally decreasing as the number of colliding users increases. Our simulation results reveal that the SUCRe protocol with $\delta=-1$ resolves $92\%$ of the two-UE collisions, $82\%$ of the five-UE collisions, and  $71\%$ of ten-UE collisions. Note that it is unlikely to have more than a handful of colliding UEs per RA pilot, except when the network is extremely overloaded.

\begin{figure} \vspace{-3mm}
        \centering
        \begin{subfigure}[t]{.95\columnwidth} \centering 
                \includegraphics[width=\columnwidth]{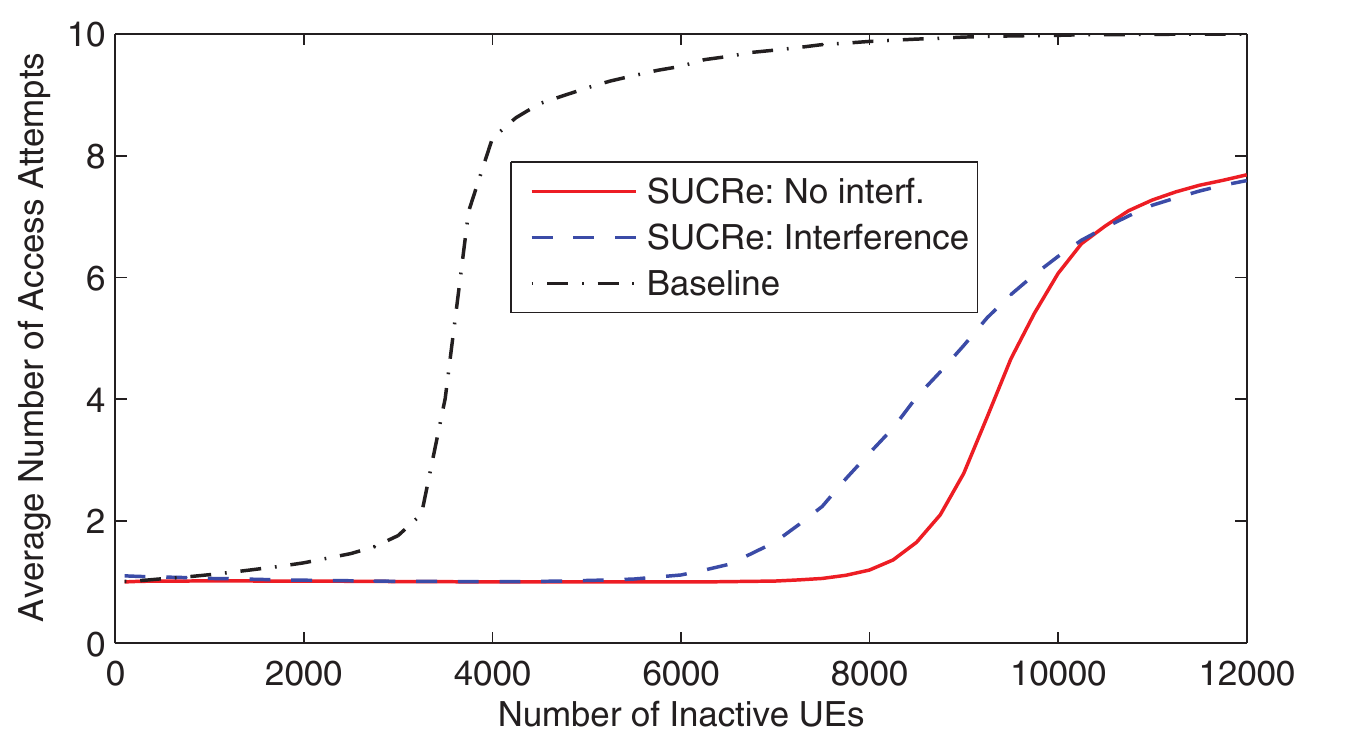} \vspace{-3mm}
                \caption{Average number of RA attempts}
                \label{figure10a}
        \end{subfigure}  \hspace{3mm}
        \begin{subfigure}[t]{.95\columnwidth} \centering
                \includegraphics[width=\columnwidth]{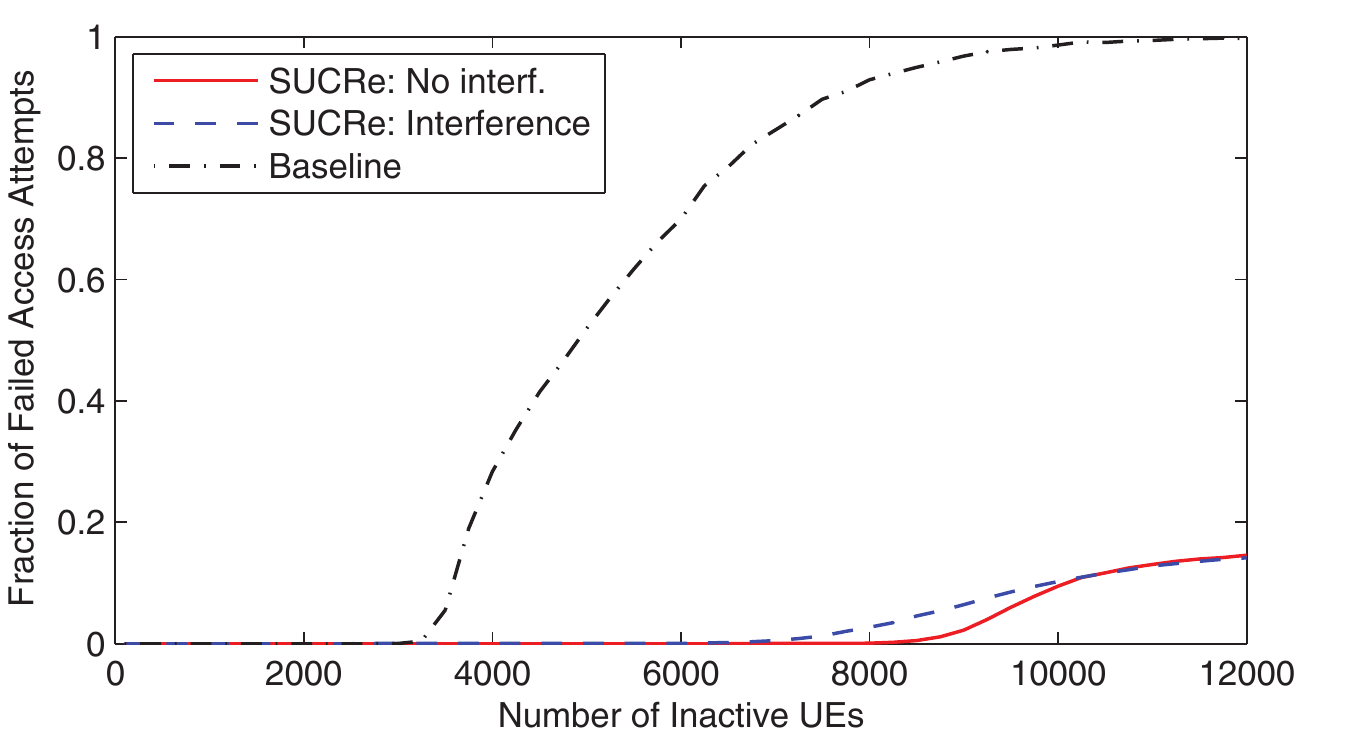} \vspace{-3mm}
                \caption{Probability of failed RA attempt (more than 10 attempts)}
                \label{figure10b} 
        \end{subfigure} 
        \caption{RA performance in a cellular network, where each UE accesses the network with $0.1 \%$ probability and sends 10 RA pilots  before giving up. The SUCRe protocol can handle substantially higher user loads, $K_0$, than conventional methods.}\label{fig:attempts} \vspace{-5mm}
\end{figure}

\subsection{Average Number of RA Attempts in Crowded Scenarios}
\label{subsec:crowded-scenario}

We have previously shown that the SUCRe protocol can resolve RA collisions, but the main purpose of an RA protocol is that every UE should be admitted to the data blocks after as few RA attempts as possible. We study this performance indicator in a scenario with $M=100$, $\tau_p=10$, and varying number of inactive UEs: $K_0 \in [100,12000]$. Each UE decides to access the network with $0.1 \%$ probability. If it is not admitted immediately, then in the upcoming blocks the UE joins the SUCRe process that runs in that block with probability $0.5$. If the UE has not succeeded after sending RA pilots in a total of $10$ SUCRe processes (including the initial one), then it stops the transmission; that is, it considers that access has been denied by the network. Note that the procedure of joining the $9$ additional SUCRe processes can be optimized according to the principles of splitting tree protocols \cite{Capetanakis1979a,Sorensen2013a}, but this optimization is outside the scope of this paper. We consider uncorrelated Rayleigh fading, cases with and without inter-cell interference, and we use the bias term $\epsilon_k = -\beta_k/ \sqrt{M} - \bar{\omega}/2$. The SUCRe protocol is compared with the same baseline protocol as before (i.e., it only handles collisions by retransmission) with the addition that the UEs make 10 RA attempts at random occasions in the same way as the SUCRe protocol.

Fig.~\ref{fig:attempts}(a) shows the average number of RA attempts that each UE makes, as a function of $K_0$, while Fig.~\ref{fig:attempts}(b) shows the fraction of UEs that fails to access the network (i.e., made 10 unsuccessful attempts). The SUCRe protocol can easily handle up to $K_0=6000$ under inter-cell interference and $K_0=8000$ when the adjacent cells are silent in the RA blocks. For larger values of $K_0$ around $0$--$15 \%$ of the UEs will fail to be admitted. Notice that at $K_0=10000$ there will on average be $K_0 \cdot 0.001 / \tau_p =1$ UE that selects each RA pilot, meaning that the network is fundamentally overloaded. Nevertheless, an astonishing $90\%$ of the UEs can still access the network successfully, which matches well with the $90\%$ probability of resolving collisions observed in Fig.~\ref{fig:cellular-simulation}. This behavior remains also for $K_0>10000$.
In contrast, the baseline protocol requires more retransmissions when $K_0 < 3000$ and as $K_0$ increases in the range $K_0>3000$ the RA functionality gradually breaks down; at $K_0=10000$ only $1.5 \%$ of the UEs are successful in their RA attempts.

\section{Conclusion} \label{sec:conclusion}

The pilot sequences are precious resources in Massive MIMO since they enable the BS to separate the UEs in the spatial domain. In future urban scenarios, the number of UEs that resides in a cell is much larger than the number of available pilots, thus the pilots need to be temporally allocated only to the UEs that have data to transmit or receive. The proposed SUCRe random access protocol provides an efficient way for UEs to request pilots for data transmission, and is well-suited for beyond-LTE Massive MIMO systems and crowded urban deployment scenarios. The protocol exploits the channel hardening and favorable propagation properties to enable distributed collision detection and resolution at the UEs, where the contender with the strongest signal gain is the one being admitted. The numerical results demonstrate that the SUCRe protocol can resolve around $90\%$ of all collisions and that it is robust to inter-cell interference and choice of channel distribution. The protocol does not break down in overloaded situations, where more UEs request pilots than there are RA resources, but continues to admit a subset of the accessing UEs.

\appendices

\section{Some Useful Results}
\label{app:useful}

\begin{lemma}[\S 8.328.2 in \cite{Gradshteyn1980a}] The Gamma function satisfies \label{lemma:gamma-property}
\begin{equation} \label{eq:gamma-property}
\frac{\Gamma \left( M + \frac{1}{2} \right) }{ \sqrt{M} \Gamma \left( M \right) } \rightarrow 1 \quad \textrm{as} \quad M \rightarrow \infty.
\end{equation}
\end{lemma}

\begin{lemma} \label{lemma:integral-result}
For any non-negative integer $m$ and real-valued $A$ and $B$, we have
\begin{equation}
\begin{split}
 & \int_{0}^{\infty} x^{m} e^{- \left( x A - B \right)^2 } dx \\  & 
 = \begin{cases} \sum\limits_{n=0}^{m} \!\! { m \choose n } \! \frac{B^{m-n}}{A^{m+1}} \frac{  \Gamma \left( \frac{n+1}{2}  \right) \!+\! (-1)^n \gamma( \frac{n+1}{2}  , B^2 ) }{2}, &B\geq 0, \\
  \sum\limits_{n=0}^{m} \!\! { m \choose n } \! \frac{B^{m-n}}{A^{m+1}} \frac{  \Gamma \left( \frac{n+1}{2}  \right) - \gamma( \frac{n+1}{2}  , B^2 ) }{2}, & B < 0.
  \end{cases}
   \end{split}
\end{equation}
\end{lemma}
\begin{IEEEproof}
By the change of variable $ x= \frac{\tilde{x}+B}{A} $ we obtain
\begin{align} \label{eq:lemma-integral-proof1}
\notag & \int_{0}^{\infty} x^{m} e^{- \left( x A - B \right)^2 } dx =\int_{-B}^{\infty} \left( \frac{\tilde{x}+B}{A}  \right)^{m} \!\!  \frac{e^{-\tilde{x}^2 }}{A}  d\tilde{x} \\
 & =  \sum_{n=0}^{m} \!\! { m \choose n } \! \frac{B^{m-n}}{A^{m+1}} \int_{-B}^{\infty} \tilde{x}^{n}  e^{-\tilde{x}^2 }  d\tilde{x} ,
\end{align}
where the second equality follows from applying the binomial formula to $( \tilde{x}+B )^{m}$. If $B \geq 0$,
the remaining integral is computed as
\begin{align} \notag
\int_{-B}^{\infty} \tilde{x}^{n}  e^{-\tilde{x}^2 }  d\tilde{x}  &
=  \int_{0}^{\infty} \tilde{x}^{n}  e^{-\tilde{x}^2 }  d\tilde{x} + (-1)^{n}  \!\! \int_{0}^{B} \tilde{x}^{n}  e^{-\tilde{x}^2 }  d\tilde{x}  \\ &
= \frac{  \Gamma \left( \frac{n+1}{2}  \right) \!+\! (-1)^n \gamma( \frac{n+1}{2}  , B^2 ) }{2} \label{eq:lemma-integral-proof2}
\end{align}
by making the variable substitution $y = \tilde{x}^2$ and identifying incomplete gamma functions.
Similarly, if $B < 0$ we have
\begin{align} \notag
\int_{-B}^{\infty} \tilde{x}^{n}  e^{-\tilde{x}^2 }  d\tilde{x}  &
=  \int_{0}^{\infty} \tilde{x}^{n}  e^{-\tilde{x}^2 }  d\tilde{x} - \!\! \int_{0}^{-B} \tilde{x}^{n}  e^{-\tilde{x}^2 }  d\tilde{x}  \\ &
= \frac{  \Gamma \left( \frac{n+1}{2}  \right) - \gamma( \frac{n+1}{2}  , B^2 ) }{2}. \label{eq:lemma-integral-proof3}
\end{align}
Substituting \eqref{eq:lemma-integral-proof2} or \eqref{eq:lemma-integral-proof3} into \eqref{eq:lemma-integral-proof1} yield the final result.
\end{IEEEproof}

\section{Collection of Proofs}
\label{app:collection-proofs}

\textbf{Proof of Lemma \ref{lemma:distribution_z_k}:}
Suppose for a moment that $\mathcal{S}_t$ is known, then the MMSE estimator of $ \vect{h}_k$ from the observation $\vect{y}_t$ at BS~$0$ is \cite{Kay1993a}
\begin{equation}
\hat{\vect{h}}_{k,\mathrm{MMSE}} = \frac{\sqrt{\rho_k \tau_p} \beta_k}{\alpha_t + \sigma^2 } \vect{y}.
\end{equation}
The true channel can be expressed as $\vect{h}_k = \hat{\vect{h}}_{k,\mathrm{MMSE}} + \vect{e}_k$, where the estimate
\begin{equation}
\hat{\vect{h}}_{k,\mathrm{MMSE}} \sim \mathcal{CN} \left(\vect{0}, \frac{\rho_k \tau_p \beta_k^2}{\alpha_t  + \sigma^2} \vect{I}_M \right)
\end{equation}
is independent from the estimation error
\begin{equation}
\vect{e}_k \sim \mathcal{CN} \left(\vect{0}, \left( \beta_k - \frac{\rho_k \tau_p \beta_k^2}{\alpha_t  + \sigma^2}  \right) \vect{I}_M \right).
\end{equation}
The BS does not know $\mathcal{S}_t$, but we notice that
\begin{equation}
\frac{\hat{\vect{h}}_{k,\mathrm{MMSE}}^* }{\| \hat{\vect{h}}_{k,\mathrm{MMSE}}  \|} = \frac{\vect{y}^*}{\| \vect{y} \|},
\end{equation}
thus the received signal in \eqref{eq:received-user-k-scalar} can be rewritten as
\begin{align} \notag
z_k &= \sqrt{q \tau_p}  \vect{h}_k^{\Ttran} \frac{\hat{\vect{h}}_{k,\mathrm{MMSE}}^* }{\| \hat{\vect{h}}_{k,\mathrm{MMSE}}  \|} + \boldsymbol{\upsilon}_k^{\Ttran} \frac{\boldsymbol{\phi}_t^*}{\| \boldsymbol{\phi}_t \|} + \eta_k \\ &  \notag
= \underbrace{\sqrt{q \tau_p} \| \hat{\vect{h}}_{k,\mathrm{MMSE}}  \|}_{=g_k} \\ &
+  \underbrace{\sqrt{q \tau_p} \frac{\vect{e}_k^{\Ttran} \hat{\vect{h}}_{k,\mathrm{MMSE}}^* }{\| \hat{\vect{h}}_{k,\mathrm{MMSE}}  \|} + \boldsymbol{\upsilon}_k^{\Ttran} \frac{\boldsymbol{\phi}_t^*}{\| \boldsymbol{\phi}_t \|} + \eta_k}_{= \nu_k},
\end{align}
where we call the two terms $g_k$ and $\nu_k$.
We notice that 
\begin{equation}
\frac{ \vect{e}_k^{\Ttran}\hat{\vect{h}}_{k,\mathrm{MMSE}}^* }{\| \hat{\vect{h}}_{k,\mathrm{MMSE}}  \|} \sim \mathcal{CN} \left( 0, \beta_k - \frac{\rho_k \tau_p \beta_k^2}{\alpha_t  + \sigma^2}  \right)
\end{equation}
since $\vect{e}_k$ is independent from the channel estimate and $\frac{\hat{\vect{h}}_{k,\mathrm{MMSE}} }{\| \hat{\vect{h}}_{k,\mathrm{MMSE}}  \|}$ is uniformly distributed over the unit sphere in $\mathbb{C}^M$. Hence, $g_k$ and $\nu_k$ are independent random variables. In addition, $\nu_k$ is the sum of three independent complex Gaussian variables which have zero mean and the total variance as stated in the lemma. Finally, we notice that $g_k^2$ is the sum of squares of $2M$ independent Gaussian variables with zero mean and variance $\frac{1}{2} \frac{\rho_k q \beta_k^2 \tau_p^2}{\alpha_t  + \sigma^2}$, thus $g_k$ has a scaled chi-distribution with $2M$ degrees of freedom as stated in the lemma.

\vspace{3mm}

\textbf{Proof of Theorem \ref{theorem:ML-estimate-alpha}:}  The ML estimator is defined as
\begin{equation} \label{eq:MLE}
\hat{\alpha}_{t,k}^\star = \argmax{\alpha} \quad f \left( z_{k,\Re}, z_{k,\Im} | \alpha \right)
\end{equation}
where $f \left( z_{k,\Re}, z_{k,\Im} | \alpha \right)$ is the joint PDF. Since the UE knows that $\rho_k   \beta_k \tau_p$, it is sufficient to search for $\alpha \geq \rho_k   \beta_k \tau_p$. Notice that $z_{k,\Re}=g_k + \Re(\nu_k)$ and $z_{k,\Im}= \Im(\nu_k)$ are independent since $\nu_k \sim \mathcal{CN}(0,\lambda_2)$ has independent real and imaginary parts that are distributed as $\mathcal{N}(0,\lambda_2/2)$. Hence,
\begin{equation} \label{eq:f1f2}
f \left( z_{k,\Re}, z_{k,\Im} | \alpha \right) = f_1 \left( z_{k,\Re} | \alpha \right) f_2 \left( z_{k,\Im} | \alpha \right) 
\end{equation}
where $f_2  \left( z_{k,\Im} | \alpha \right)$ in \eqref{eq:f2-pdf} is the PDF of $\Im(\nu_k)$. It remains to compute the PDF $f_1 \left( z_{k,\Re} | \alpha \right)$ of $z_{k,\Re}$, which is a convolution of the PDFs of $g_k$ and $\Re(\nu_k)$:
\begin{align} \notag
&f_1 \left( z_{k,\Re} | \alpha \right) = \int_{0}^{\infty} 
\frac{2 x^{2M-1} e^{-x^2/\lambda_1}}{\Gamma(M) \lambda_1^M}
\frac{e^{- ( z_{k,\Re} - x  )^2 / \lambda_2 }}{\sqrt{\pi \lambda_2 }}  dx \\ \notag
&= \frac{2 e^{- z_{k,\Re}^2 / \lambda_2 } }{\Gamma(M) \lambda_1^M \sqrt{\pi \lambda_2 }} \int_{0}^{\infty} x^{2M-1} e^{-x^2 ( \frac{1}{\lambda_1} + \frac{1}{\lambda_2}) + x \frac{2  z_{k,\Re}}{\lambda_2}  } dx \\
&= \frac{2 e^{- \frac{(z_{k,\Re})^2}{\lambda_2}  \left( 1 - \frac{\lambda_1}{\lambda_1 + \lambda_2}   \right)} }{\Gamma(M) \lambda_1^M \sqrt{\pi \lambda_2 }} \!\!  \int_{0}^{\infty} x^{2M-1} e^{- \left( x A - B \right)^2 } dx
 \label{eq:first-step}
\end{align}
where $A = \sqrt{ \frac{1}{\lambda_1} + \frac{1}{\lambda_2} } $ and $B = \frac{z_{k,\Re}}{\lambda_2 A} $. The final expression for $f_1$ in \eqref{eq:f1f2} is obtained by computing the integral in  \eqref{eq:first-step} using Lemma \ref{lemma:integral-result} with $m=2M-1$.

\vspace{3mm}

\textbf{Proof of Theorem \ref{theorem:retransmission-probability}:}  
The probability of repeating the pilot in Step 3, based on \eqref{eq:retransmission-rule}, is 
\begin{equation}
\Pr \{ \mathcal{R}_{k} \} = \Pr \left\{ 2 \rho_k \beta_k \tau_p > \hat{\alpha}_{t,k}^{\text{approx2}} + 2\epsilon_k  \right\}.
\end{equation}
Notice that whenever $\epsilon_k < \rho_k   \beta_k \tau_p/2$, $\hat{\alpha}_{t,k}^{\text{approx2}} = \rho_k   \beta_k \tau_p$ will always lead to pilot repetition, thus we can neglect the maximum-operator in  \eqref{eq:alpha-approx-est2} and write
\begin{align} \notag
\Pr \{ \mathcal{R}_{k} \} &= \Pr \left\{ 2\rho_k \beta_k \tau_p > C_M^2
\frac{ q \rho_k   \beta_k^2 \tau_p^2  }{ \left( \Re ( z_k ) \right)^2 } - \sigma^2   + 2\epsilon_k  \right\} \\
&=  \Pr \left\{ \left( \Re ( z_k ) \right)^2 > \zeta_k  \right\}
\label{eq:prob-derivation1}
\end{align}
where we use the notation $C_M = \Gamma \left( M \!+\! \frac{1}{2} \right) / \Gamma \left( M \right) $ and the second equality follows from rearranging the term and identifying $\zeta_k$ from \eqref{eq:zeta-parameter}. Since $ \Re ( z_k )$ can be negative, we rewrite \eqref{eq:prob-derivation1} as
\begin{equation}
\begin{split}
&\Pr \{ \mathcal{R}_{k} \}  =  \Pr \left\{  \Re ( z_k )  > \sqrt{\zeta_k}  \right\} +  \Pr \left\{  \Re ( z_k )  < -\sqrt{\zeta_k}  \right\} \\
&= 1 -  \Pr \left\{  \Re ( z_k )  \leq \sqrt{\zeta_k}  \right\} +  \Pr \left\{  \Re ( z_k )  \leq -\sqrt{\zeta_k}  \right\} .
\end{split}
\end{equation}
which utilizes the fact that $\Pr \{  \Re ( z_k )  < -\sqrt{\zeta_k}  \} = \Pr \{  \Re ( z_k )  \leq -\sqrt{\zeta_k}  \}$.
It remains to compute the CDF $\Pr \left\{  \Re ( z_k )  \leq  d \right\}$ for an arbitrary $d$, which is the convolution of the CDF of $g_k$ and the PDF of $\Re(\nu_k)$:
\begin{align} \notag
&\Pr \{ \Re ( z_k )  \leq b \}  = \int_{0}^{\infty} \frac{ \gamma \left(  M, \frac{x^2}{2\lambda_1} \right)
}{\Gamma(M)}
\frac{e^{- ( b  - x  )^2 / \lambda_2 }}{\sqrt{\pi \lambda_2 }}  dx \\ \notag
&= \int_{0}^{\infty} \frac{e^{- ( b  - x  )^2 / \lambda_2 }}{\sqrt{\pi \lambda_2 }}  dx \\ &
- \int_{0}^{\infty}  e^{-x^2/\lambda_1} \sum_{k=0}^{M-1} \frac{x^{2k}}{\lambda_1^k \Gamma(k+1)} 
\frac{e^{- ( b  - x  )^2 / \lambda_2 }}{\sqrt{\pi \lambda_2 }}  dx  \label{eq:prob-derivation2}
\end{align}
by utilizing the definition of the incomplete gamma function (see Footnote \ref{footnote:incomplete-gamma}). The first integral in \eqref{eq:prob-derivation2} is over a Gaussian PDF and identified as $Q(-b\sqrt{2/\lambda_2})$. The second integral can be rewritten as
\begin{align} \notag
&\sum_{k=0}^{M-1} \frac{e^{- b^2/\lambda_2 } }{\Gamma(k+1)  \lambda_1^k  \sqrt{\pi \lambda_2 }} 
   \int_{0}^{\infty}  x^{2k} e^{-x^2 ( \frac{1}{\lambda_1} + \frac{1}{\lambda_2}) + x \frac{2 b}{\lambda_2}  }  dx  \\
  &= \sum_{k=0}^{M-1} \frac{ e^{- \frac{b^2}{\lambda_2}  \left( 1 - \frac{\lambda_1}{\lambda_1 + \lambda_2}   \right)} }{ \Gamma(k+1) \lambda_1^k \sqrt{\pi \lambda_2 } } 
   \int_{0}^{\infty}  x^{2k} e^{-(xA-B)^2  }  dx \label{eq:prob-derivation3}
\end{align}
where $A = \sqrt{ \frac{1}{\lambda_1} + \frac{1}{\lambda_2} } $ and $B = \frac{b}{\lambda_2 A} $.
The final expression in \eqref{eq:cdf-zre} is obtained from \eqref{eq:prob-derivation2}--\eqref{eq:prob-derivation3} by computing the remaining integral using Lemma \ref{lemma:integral-result} with $m=2k$.

\vspace{3mm}

\textbf{Proof of Corollary \ref{cor:asymptotic-CDFs}:}  
The Lindeberg-L\'evy central limit theorem implies that $g_k$ converges to 
$\mathcal{N}(  \sqrt{ \lambda_1 } C_M, \lambda_1^2(M- C_M^2  ) )$ in distribution as $M \rightarrow \infty$, where $C_M = \Gamma \left( M \!+\! \frac{1}{2} \right) / \Gamma \left( M \right)$. Since $ \Re ( z_k ) = g_k +  \Re ( \nu_k )$ converges to the sum of two independent Gaussian variables (recall Lemma \ref{lemma:distribution_z_k}), the CDF is obtained by \eqref{eq:asympotic-CDF}.

Notice that \eqref{eq:probability-retransmit} can be expressed as
\begin{equation}
\Pr \{ \mathcal{R}_{k} \} = 1 + \Pr \{ \Re ( z_k )   > \sqrt{\zeta_k} \}) - \Pr \{ \Re ( z_k )   > - \sqrt{\zeta_k} \}.
\end{equation}
Since $\zeta_k$ in \eqref{eq:zeta-parameter} is positive, $\Pr \{ \Re ( z_k )   > - \sqrt{\zeta_k} \} \rightarrow Q(-\infty) = 1$ as $M \rightarrow \infty$.
Similarly, we notice that
\begin{equation} 
\Pr \{ \Re ( z_k )   > \sqrt{\zeta_k} \}) \rightarrow \begin{cases} 1, & \zeta_k<C_M^2 \lambda_1, \\ 1/2, & \zeta_k=C_M^2 \lambda_1, \\ 0, & \zeta_k>C_M^2 \lambda_1. \end{cases}
\end{equation}
The asymptotic probabilities in \eqref{eq:asymptotic-retransmission-prob} follows directly from these results.

\bibliographystyle{IEEEbib}
\bibliography{IEEEabrv,refs}

\end{document}